\documentclass[10pt,journal,compsoc]{IEEEtran}
\usepackage{amsmath,amsfonts}
\usepackage{algorithmic}
\usepackage{graphicx}
\usepackage{textcomp}

\usepackage{xspace}
\usepackage{comment}
\usepackage{array}

\usepackage[
 svgnames,
 table,
]{xcolor}

\usepackage[
 colorlinks,
 urlcolor = NavyBlue,
 citecolor = NavyBlue,
 linkcolor = NavyBlue,
 ]{hyperref}
\usepackage{tikz}
\usetikzlibrary{automata, positioning, arrows.meta}
\usepackage{wrapfig}
\usepackage{caption}
\usepackage{subcaption}
\usepackage{multirow}
\usepackage{booktabs}
\usepackage{adjustbox}
\usepackage{diagbox}

\usepackage[numbers]{natbib}
\usepackage{url}

\usepackage{cleveref}
\crefname{equation}{Eq.}{Eqs.}
\Crefname{equation}{Equation}{Equations}
\crefname{proposition}{Prop.}{Props.}
\Crefname{proposition}{Proposition}{Propositions}
\crefname{lemma}{Lemma}{Lemmata}
\Crefname{lemma}{Lemma}{Lemmata}
\crefname{listing}{Listing}{Listings}
\Crefname{listing}{Listing}{Listings}
\crefname{definition}{Def.}{Definitions}
\Crefname{definition}{Definition}{Definitions}
\crefname{theorem}{Thm.}{Theorems}
\Crefname{theorem}{Theorem}{Theorems}
\crefname{figure}{Fig.}{Figs.}
\Crefname{figure}{Figure}{Figures}
\crefname{page}{p.}{pages}
\Crefname{page}{Page}{Pages}
\crefname{section}{Sect.}{Sects.}
\Crefname{section}{Section}{Sections}
\crefname{example}{Example}{Examples}
\Crefname{example}{Example}{Examples}
\crefname{table}{Tbl.}{Tables}
\Crefname{table}{Table}{Tables}

\definecolor{Gray}{gray}{0.92}
\newcolumntype{a}{>{\columncolor{Gray}}c}
\newcolumntype{b}{>{\columncolor{Gray}}r}
\newcolumntype{R}{@{\extracolsep{15pt}}c@{\extracolsep{0pt}}}
\newcolumntype{A}{@{\extracolsep{15pt}}>{\columncolor{Gray}}c@{\extracolsep{0pt}}}

 \usepackage[draft,silent]{fixme}
\fxsetup{theme=color,mode=multiuser}
\FXRegisterAuthor{pk}{PK}{Paul}
\FXRegisterAuthor{ebj}{EBJ}{Einar}
\FXRegisterAuthor{ap}{AP}{Andrea}
\usepackage[clean]{revdiff}

\usetikzlibrary{positioning,calc,arrows,shapes,automata,fit,ducks}

\newtheorem{definition}{Definition}
\newtheorem{example}{Example}
\newcommand{\name}[1]{\textsc{\small #1}\xspace}

\def\BibTeX{{\rm B\kern-.05em{\sc i\kern-.025em b}\kern-.08em
    T\kern-.1667em\lower.7ex\hbox{E}\kern-.125emX}}

\newcommand{\etal}{\emph{et al.}\xspace}

\DeclareMathOperator{\head}{head}
\DeclareMathOperator{\tail}{tail}

\DeclareMathOperator{\loops}{loops}

\DeclareMathOperator{\var}{var}

\newcommand{\unit}[1]{\,\mathrm{#1}}
\makeatletter
\newcommand\notsotiny{\@setfontsize\notsotiny\@vipt\@viipt}
\makeatother

\newcommand*\rot{\rotatebox{90}}
\newcolumntype{s}{>{\columncolor[HTML]{AAACED}} p{3cm}}

\providecommand{\doi}[1]{\url{https://doi.org/#1}}
\IEEEpubid{\doi{10.1109/TSE.2026.3679253}\$00.00~\copyright~2026 IEEE}

\begin{document}

\title{Automata Learning versus Process Mining:\\ The Case for User Journeys} 

\author{Paul Kobialka, Andrea Pferscher, Bernhard K. Aichernig, Einar Broch Johnsen, Silvia Lizeth Tapia Tarifa}

\markboth{IEEE Trans. Software Eng. (\MakeLowercase{\emph{preprint, to appear}} 2026)}%
{Kobialka \MakeLowercase{\emph{et al.}}: Automata Learning versus Process Mining: The Case for User Journeys}

\IEEEtitleabstractindextext{%
  
\begin{abstract}
  With the servitization of business, understanding how users
  experience services becomes a crucial success factor for
  companies. Therefore, there is a need to include feedback from user
  experiences in the software engineering process. Behavioral models
  of user journeys, describing how users experience their interaction
  with a service, can provide insights and potentially improve
  services. In this paper, we investigate techniques that allow the
  automatic generation of behavioral models from user interactions
  with a service, recorded in an event log.  We first compare two
  established techniques that generate behavioral models from a given
  event log: automata learning and process mining. Afterward, we
  present a novel, hybrid method that combines both automata
  learning and  process mining methods to overcome their
  limitations. For the existing techniques, we present methods to
  learn models of user journeys and evaluate the accuracy of the
  resulting models.  We then compare these techniques with our novel
  method for the automatic extraction of user journey models from the
  event logs of digital services.  We assess the practical
  applicability of all techniques by evaluating real-world
  applications. Our results show that process mining techniques rely
  on expert knowledge, while automata learning techniques depend on
  the distribution of events in the given event log. We further show
  that the proposed hybrid technique combines the strengths of both
  process mining and automata learning, automatically selecting the
  best method and parameter settings for a given event log to learn
  very accurate models.
\end{abstract}
\begin{IEEEkeywords}
passive automata learning, process discovery, model inference, model learning, customer journeys, AALpy
\end{IEEEkeywords}}

\maketitle
\IEEEpubidadjcol
\IEEEdisplaynontitleabstractindextext

% \copyrightnotice

% Remove this page numbers later in submission
% \thispagestyle{plain}
% \pagestyle{plain}
% page limit 18 pages for all text and figures (plus 4 pages for references)
\section{Introduction}

Software products increasingly rely on service-oriented business
models~\cite{vandermerweServitizationBusinessAdding1988}, making user
satisfaction
% \pknote{when interacting with services}
a key factor in a business' financial
success~\cite{fornellCustomerSatisfactionStock2006}.  \emph{User (or
  customer) journeys}~\cite{ROSENBAUM2017143} are models that describe
how users experience a service.  \rnew{These models have been highly
  successful in explaining and improving
  services~\cite{halvorsrudImprovingServiceQuality2016}}, but their
construction and analysis typically require significant manual
effort~\cite{halvorsrudSmartJourneyMining}. This paper investigates
the automated, data-driven construction of behavioral models of user
journeys, that are amenable to tool-driven analysis. We compare two
techniques for learning such behavioral models from the system logs of
services: \emph{automata learning} (AL)~\cite{vaandrager2017model} and
\emph{process mining} (PM)~\cite{DBLP:books/sp/Aalst16}.

For user-centric services, it is important to feed insights about user
behavior back into the continuous development process (e.g.,
\cite{pahl04ist}).  \rnew{These services are typically personalized~\cite{rust2014service}}, and
changes to the service might directly affect user behavior. For the
service provider, it is crucial to know why and in which steps of the
service they lose users.  Usually, if companies grow over time, so do
the number of users and the variations in the service.  
\rnew{Companies are confronted with evolving user expectations~\cite{rust2014service,lemon2016understanding}}; their success depends on
maintaining the quality of the constructed behavioral models.
However, user journey models are usually generated and analyzed
manually, typically with a few selected users through questionnaires,
relying heavily on domain experts~\rnew{\cite{folstad2018customer}}.
Thus, the creation of user journey
models lacks both agility and scalability to large user bases.  The
process suffers from a lack of tool support and is inherently limited
by the high amount of manual
labor~\cite{halvorsrudSmartJourneyMining}.\looseness=-1

User journeys record and analyze the time sequence of interactions
between a service provider and a user from the user's perspective.
Users interact with a service to reach some goal, e.g., to order an
item online.  The steps of the user journey, called
\emph{touchpoints}, record actions or communications between a service
provider and a user.  Touchpoints are observable and stateful, which
makes user journeys amenable to behavioral modeling.  Behavioral
models of user journeys describe user behavior using a finite-state
representation, providing insights into an underlying user journey. A
formal representation of the user behavior enables automatic analysis
of properties, e.g., checking whether a user can achieve a certain
task. In practice, the presence of such behavioral models is
limited. Therefore, this paper investigates techniques to
automatically construct behavioral models of user journeys.\looseness=-1

% The assumption that these touchpoints are observable and stateful
% allows us to formalize user journeys as behavioral models.

\begin{figure}[t]
    \centering
    \includegraphics[width=\linewidth]{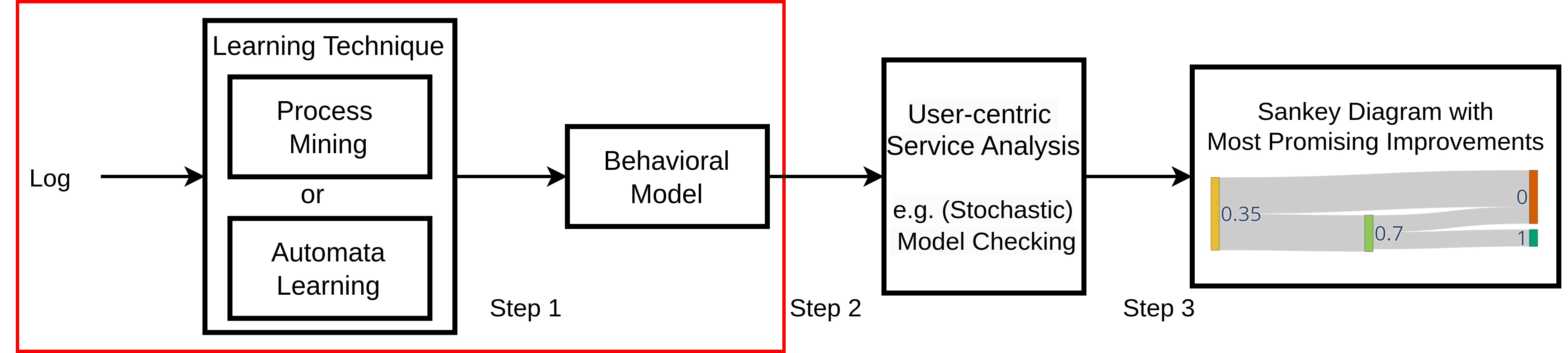}
    \caption{Three step procedure for the creation and analysis of
      user journey models. This work focuses on Step\,1, indicated by
      the red square, i.e., the automatic creation of behavioral
      models from event logs.}
    \label{fig:overview}
\end{figure}

Automatically constructed behavioral models of user journeys have
recently been used to analyze weaknesses in service offerings by
\Citet{kobialkaFM24,kobialka24sosym}; the user journeys were
formalized as transition systems, constructed from system logs using
PM and AL techniques. \Cref{fig:overview} illustrates the three-step
procedure introduced by \Citet{kobialkaFM24} to automatically generate
behavioral models from logs, enabling advanced model-based
analysis. In Step~1, we can either use AL or PM to create behavioral
models from logs. Compared to manually constructed models, the
approach enables the analysis of user journeys at a completely
different scale; system logs may span from days to years of
interaction between service and users, captured in possibly more than
thousands of traces. Pain points in the offered service are
automatically detected by (probabilistically) model checking~\cite{10.1007/BFb0058022,DBLP:reference/mc/BaierAFK18} the
behavioral model in Step~2.  Step~3 includes further visualization;
e.g., \emph{Sankey} diagrams~\cite{riehmann05infovis} can then show us
the actions most viable for improvements.
\rnew{The pipeline shown in \cref{fig:overview} demonstrates that finite transition systems enable an in-depth analysis of user behavior. Analysis techniques such as model checking can identify bottlenecks in a service. Kobialka~\etal~\cite{kobialkaFM24} also use model checking to visualize user flows in a more understandable way for service providers, via Sankey diagrams.}
% ~\cite{kobialkaWeightedGames,kobialkaEdba22,kobialka24sosym,kobialka24abs,kobialkaFM24}

In this paper, we focus on Step 1 in \cref{fig:overview} and compare
the two alternatives for constructing user journey models, using
either PM or AL techniques. Both are well-established, but stem from
different communities: \rchange{(1) \emph{process discovery} in PM utilizes
logs to generate a process model, e.g., a transition system, Petri net
or BPMN model~\cite{DBLP:books/sp/Aalst16}, and (2) \emph{passive AL}
constructs a finite automaton accepting the formal language of a given
log~\cite{DBLP:journals/iandc/Gold67}}{PM techniques aim to gain insights into business processes, while AL techniques aim to derive the most general representation of a black-box system as a behavioral model. Nevertheless, there are similarities: \emph{Process discovery} is a step in the PM pipeline that addresses the problem of finding a model representation from a given event log. \emph{Passive automata learning} has a similar goal, but focuses on finding the most general representation.}
Recent developments in PM
emphasize discovery techniques for \rnew{rich} behavioral models, e.g., Petri nets \rnew{or process trees}
that allow the formalization of advanced system characteristics such
as parallelism~\cite{van2022foundations}. 
However, user journeys do not require such a
feature-rich modeling formalism. As shown
by~\citet{kobialkaWeightedGames,kobialka2025decision}, finite transition
systems model user journeys sufficiently \rnew{and allow for rich analyses}. 
\rnew{Therefore, to enable realistic analyses, we compare PM and AL for learning behavioral models of user journeys.}\looseness=-1

\rnew{Finite transition systems enable}
the application of AL techniques, which provide well-established
algorithms~\cite{deLaHigueraRPNI,DBLP:conf/icgi/CarrascoO94,verwerFlexFringeModelingSoftware2022}
for learning a minimal finite-state representation of a given set of
system traces. However, finding the right
level of behavioral approximation defined by a generated model is
\emph{the most interesting challenge in process mining}~\cite{van2010process}. The
comparison of AL and PM techniques specifically addresses this
challenge: AL targets the most general state representation in terms
of overapproximation, whereas PM approaches an underapproximation. Our
comparison between PM and AL shows that the accuracy of the generated
model strongly depends on the underlying event log, where PM is
beneficial for sparse event logs and AL for larger, well-distributed
logs. In this paper, we propose a novel method, called \name{Hybrid},
that automatically applies either PM or AL depending on characteristics of the event log to obtain the most accurate model. \looseness-1

\smallskip

\textbf{Contributions.}  The key contributions of this paper are: (1)
a novel benchmark suite for the evaluation of user journey learning
techniques, (2) an exhaustive comparison of different AL and PM
learning setups based on the benchmark suite, (3) the novel method
\name{Hybrid} that combines established AL and PM methods to extract
behavioral user journey models in practice, (4) a practical evaluation
that includes four different real-world case studies, and (5) a
discussion of actionable insights from these experiments.

%%% Local Variables:
%%% mode: latex
%%% TeX-master: "main"
%%% End:

\section{Related Work}\label{sec:related-work}
Recent work on user journey modeling
by~\citet{halvorsrudSmartJourneyMining,halvorsrud2023involving}
emphasize the need for digital support.  In their work, the
\emph{actual journeys}, reflected as user experience in the models,
are manually discovered through interviews with
users~\cite{halvorsrudImprovingServiceQuality2016}. We here discuss
previous work on automatic model generation for user journeys, based
on PM and AL techniques. In contrast to all related work, our work
focuses on \emph{comparing} and \emph{combining} PM and AL techniques. 
To \rnew{the best of} our knowledge, we are the first to propose a combination of AL and PM techniques.

\rchange{PM-based methods}{Process discovery methods in PM} enable a data-driven approach to generate user
journey models. Positioning user journey mappings in the PM landscape,
\citet{bernardProcessMiningBased,bernardCJMexGoalorientedExploration,bernardCJMabAbstractingCustomer2018,bernardContextualBehavioralCustomer2019}
consider methods to abstract large numbers of user journeys into
compact representations, including an XML format for user
journeys~\cite{bernardProcessMiningBased}, hierarchical clustering
\cite{bernardCJMexGoalorientedExploration}, user journey maps with
different levels of granularity based on process trees
\cite{bernardCJMabAbstractingCustomer2018}, and a genetic algorithm to
build representative user journeys
\cite{bernardContextualBehavioralCustomer2019}.
\citet{harbichDiscoveringCustomerJourney} discover user journey maps
using mixtures of Markov models. %
\citet{terragniAnalyzingCustomerJourney2018,terragniOptimizingCustomerJourney2019}
use PM tools to generate models of an underlying user journey for
different user groups, and optimize towards specific key performance
indicators. 
\rold{Verbeek~\cite{verbeek2013bpi} generates transition system models for the BPI Challenge 2012~\cite{bautista2012bpic12challenge}.}
\rnew{The BPI Challenge 2012~\cite{bpi2012,bautista2012bpic12challenge} 
makes real-life event logs from a financial institution available for analysis. \citet{verbeek2013bpi} addresses the challenge by constructing multiple transition systems for different aspects of the event log, each with an adjusted representation. The results show the potential of analyzing transition systems to gain process-specific insights. We include the BPI challenge 2012 and 2017~\cite{bpi2017} among the use cases for evaluating our approach.}
\citet{kobialkaWeightedGames,kobialka2025decision,kobialka24sosym} develop
PM techniques to generate and analyze weighted games that reflect the
user experience from logs.
\rnew{This line of work demonstrates the usefulness of finite state automata for representing user journeys; hence, we use them in this paper. 
There are also process discovery techniques without an explicit model representation; e.g.,
declarative process discovery \cite{DBLP:journals/tmis/CiccioM15,DBLP:conf/caise/MaggiBA12} represents models by temporal logic formulae. These representations are not addressed in our work.}
% motivates the suitability of finite state representations in PM. In contrast to our work, this branch of work assumes a set of given constraints for inferring the underlying model. However, our work investigates techniques based on a black-box assumption of the underlying service, which overcomes the limitation of having a-priori knowledge available. 

In AL, methods for passive learning can be used for event logs. The
earliest work we know that uses AL for process models is by
\citet{cook1998discovering}, who learn a model from software
logs. \citet{DBLP:journals/fuin/EsparzaLS11} proposed an active AL
algorithm to learn automata from Petri nets. Their paper also
discusses an extension to learning from event logs that requires
comprehensive logs. \citet{agostinelli2023process} evaluate AL methods
for event logs and conclude that the active $L^*$-algorithm
of~\citet{DBLP:journals/iandc/Angluin87} is not suitable for model
generation, but passive algorithms perform well compared to the PM
algorithm \emph{Declare} of \citet{pesic07edoc}, e.g.,
RPNI~\cite{deLaHigueraRPNI}, MDL~\cite{DBLP:conf/icgi/AdriaansJ06},
and EDSM~\cite{DBLP:conf/icgi/LangPP98}. In contrast to this work, they assume that only traces of specific length should be included in the model. \citet{kobialkaFM24}
also apply AL using the complete event log, but unlike this work, they learn stochastic weighted games.
% \rnew{or use the learned models for process explanations with counterfactuals~\cite{kobialka2025Counterfactual}}. 
In a similar work, \citet{DBLP:conf/birthday/JohnsenKPT25} show the capability of AL for dealing with large data sets, creating timed stochastic games for music streaming behavior.
\citet{wieman2017experience} report on experiences from an industrial
case study using passive AL
% in a payment company
to mine models of the developed software tool from test logs, and
discuss practical applications of AL to improve software
development. A similar empirical approach could be applied using
learned user journeys to improve services.

\rnew{Akin to user journeys, \emph{software product lines} (or \emph{software process lines}) define a base process from which multiple variations are derived.
%  We review work using AL and PM for learning models of software process lines.
There are both PM~\cite{blum2015software,DBLP:conf/ispw/RubinGAKDS07,DBLP:conf/centeris/CheminguiGMSG19} and AL~\cite{damasceno2019learning} techniques to generate and analyze models of the underlying process. 
The PM techniques for software process lines have a similar setup as for user journey mining, i.e.~using process discovery techniques to generate a model from a process log. However, the AL approaches differ:
%  Blum~\etal~\cite{blum2015software} present a designated software process line discovery algorithm which differentiates between the base process, variations, and noise.
Damasceno~\etal~\cite{damasceno2019learning} extend AL techniques for software product lines by learning featured finite state machines over collections of product models. Software product lines are  not addressed in our current work.}

\rnew{No related work compares PM and AL in terms of learning behavioral models of user journeys. Our paper addresses this gap by focusing on logs of varying sizes as observed in practice.
No related work investigates whether learned behavioral models over- or under-approximate the given event log, and which parameter settings are adequate for user journeys.
Our work examines how to set model parameters for PM and AL for sparse logs and proposes a novel \name{HYBRID} technique which applies PM on small and AL on large logs. We summarize the lessons learned from our experiments into practical insights for practitioners.}
%%% Local Variables:
%%% mode: latex
%%% TeX-master: "main"
%%% End:

\section{Preliminaries}
\label{sec:preliminaries}

We briefly introduce the required techniques from PM and AL, after
providing some basic definitions.
Given a finite set $A$ of events, a \emph{trace} $\sigma \in A^*$ is a
finite \rnew{sequential} sequence over $A$\rnew{, possibly with duplicates}. An \emph{event log} $L$ is a multiset of
traces in $A^*$.
%we denote by $A^*$ the set of possibly infinite sequences (or traces) over the events in $A$. A multiset $M$ over $A$ maps every element of $A$ to a natural number. Let $\supp(M)$ denote the \emph{support} of $M$, i.e., the set of elements $x\in A$ such that $M(x)>0$, and $\mathcal{B}(A)$ the set of multisets over $A$.
% \begin{definition}[Event logs]\label{def:eventlog}
%   Let $A$ be a set of events.  An \emph{event log}
%   $L \in \mathcal{B}(A^*)$ is a multiset of traces over $A$.
% \end{definition}
Let $\epsilon$ denote the empty trace. For a trace
$\sigma = \langle a_0, \dots, a_{n} \rangle$ with length
$\lvert \sigma \rvert = n + 1$, $\sigma_i$ denotes the $i^{\text{th}}$ event of $\sigma$
(so $\sigma_i=a_i$) and $\sigma_{i:j}$ the subtrace
$\langle a_i, \dots, a_{j-1} \rangle$.  If $i \geq j$, then
$\sigma_{i:j}=\epsilon$.  We abbreviate $\sigma_{i:\lvert \sigma \rvert}$ by
$\sigma_{i:}$ and $\sigma_{0:i}$ by $\sigma_{:i}$.\looseness-1

% \subsection{Transition Systems}
% \label{sec:ts}
% Given a set of events $A$, we denote by $A^*$ the set of possibly
% infinite sequences (or traces) over the events in $A$. A multiset
% $M$ over $A$ maps every element of $A$ to a natural number. Let
% $\supp(M)$ denote the \emph{support} of $M$, i.e., the set of
% elements $x\in A$ such that $M(x)>0$, and $\mathcal{B}(A)$ the set
% of multisets over $A$.  We define an event log $L$ to be a multiset
% of traces over $A$, $L \in \mathcal{B}(A^*)$.

% \begin{definition}[Event logs]\label{def:eventlog}
%   Let $A$ be a set of events.  An \emph{event log}
%   $L \in \mathcal{B}(A^*)$ is a multiset of traces over $A$.
% \end{definition}

% Let $\epsilon$ denote the empty trace.  For a trace
% $\sigma = \langle t_0, \dots, t_{n} \rangle$, let $\sigma_i$ denote
% the $i$-th event of $\sigma$ (so $\sigma_i=t_i$) and $\sigma_{i:j}$
% the subtrace $\langle t_i, \dots, t_{j-1} \rangle$.  If $i \geq j$,
% this subtrace is $\epsilon$.  Let $\sigma_{i:}$ abbreviate the
% subtrace $\sigma_{i:|\sigma|}$ and $\sigma_{:i}$ the subtrace
% $\sigma_{0:i}$.

\begin{figure}
\scalebox{0.92}{
  \begin{tikzpicture}[
    % Global styles for the tikzpicture
    % -{Stealth[length=2mm]} defines the arrow tip style
    % shorten >=1pt shortens arrows slightly to avoid overlapping with node borders
    % auto places labels automatically
    % node distance=3cm sets default distance between nodes
    % every state/.style defines appearance of all 'state' nodes
    -{Stealth[length=2mm]},
    shorten >=1pt,
    auto,
    node distance=2cm,
    font=\scriptsize,
    every state/.style={draw=black, thick,, circle, minimum size=0.8cm, inner sep=0.3pt, font=\scriptsize}
]

    % Define the states (nodes)
    % (q1) is the node name, {State 1} is the text label displayed inside the node
    \node[state, initial,initial text=, text width=1.15cm, align=center,font=\sffamily\scriptsize] (q0) {start};
    % Position q2 to the right of q1
    \node[state] (q1) [right = 1.2cm of q0, text width=1.15cm,align=center,font=\sffamily\scriptsize] {Q1 answered};
    \node[state] (q2) [below = 0.75cm of q1, text width=1.15cm,,align=center,font=\sffamily\scriptsize] {Q2 answered};
    \node[state, accepting] (q3) [above right = 0.25cm and 0.7cm of q2, text width=1.15cm,,align=center,font=\sffamily\scriptsize] {quit};
    \node[state] (q4) [below right = 0.25cm and 0.7cm of q2, text width=1.15cm,,align=center,font=\sffamily\scriptsize] {results};
    \node[state,accepting] (q5) [right = 1cm of q4, text width=1.15cm,,align=center,font=\sffamily\scriptsize] {accepted};

    \draw (q0) edge node[font=\itshape\scriptsize] {answer Q1} (q1);
    
    % Draw a loop on q2
    % [loop above] specifies the direction of the loop
    \draw (q1) edge [loop above, looseness=4,font=\itshape\scriptsize] node {re-work Q1} (q1);
    \draw (q1) edge[font=\itshape\scriptsize] node[left] {answer Q2} (q2); 
    \draw (q2) edge[font=\itshape\scriptsize] node {quit} (q3);
    \draw (q2) edge[font=\itshape\scriptsize] node[yshift=1mm, xshift=-2mm] {grading} (q4);
    \draw (q4) edge[font=\itshape\scriptsize, text width=2cm, align=center] node {accept \\ grading} (q5);

\end{tikzpicture}}
  \caption{\rnew{Example of a transition system.}}\label{fig:ts}
\end{figure}
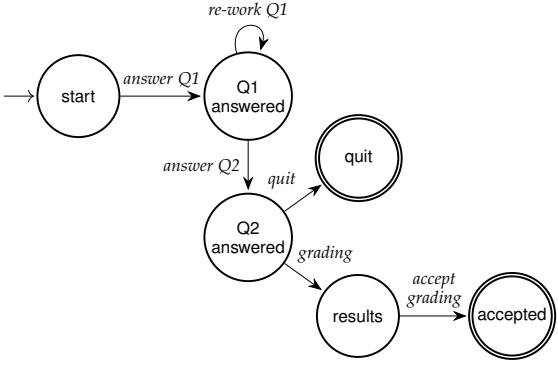

A \emph{transition system}
(TS)~\cite{nielsen1992elementary,plotkinStructuralApproachOperational}
is a tuple $\mathit{TS}=(\Gamma, A, E, I, T)$, where
\begin{itemize}
  \item $\Gamma$ is a finite set of states,
  \item $A$ a finite set of events,
  \item $E \subseteq \Gamma \times A \times \Gamma$ the finite
set of transitions, 
  \item  $I \subseteq \Gamma$ the finite
  set of initial states, and
  \item $T \subset \Gamma$ the finite set of final states.
\end{itemize}

A \emph{run} in $\mathit{TS}$ is an alternating sequence of states and
events $\langle \mathsf{s_0}, a_0, \mathsf{s_1}, \dots\rangle$ such that
$(\mathsf{s_i}, a_i, \mathsf{s_{i+1}}) \in E$, starting from an initial state
$s_0 \in I$.  Let $R$ denote the set of all runs in $\mathit{TS}$ and
$\pi(r) = \langle a_0, a_1, \ldots \rangle$ the projection of a run
$r \in R$ to its corresponding trace of events. Let
$\mathcal{L}(\mathit{TS})$ denote the language induced by
$\mathit{TS}$, defined by all runs that end in a final state, i.e.,
$\mathcal{L}(\mathit{TS}) = \{ \pi(r) \mid r \in R \land \mathsf{s_n} \in T
\}$. We call a trace $\sigma \in A^*$ \emph{positive} iff
$\sigma \in \mathcal{L}(\mathit{TS})$, otherwise it is
\emph{negative}. $\mathit{TS}$ is \emph{deterministic} if
$(\mathsf{s}, a, \mathsf{s'}), (\mathsf{s}, a, \mathsf{s''}) \in E \to \mathsf{s'} = \mathsf{s''}$. \rnew{In the sequel, we
  assume a TS to be non-deterministic.}  \rnew{Runs and traces
  highlight the difference between a transition system and an event
  log: in a run on the transition system both the events and the
  visited states are known, while the corresponding trace in the log
  only records events.}
% \rnew{We differentiate between runs and traces to highlight the
% difference between the run on a transition system, where the visited
% states are known, and the corresponding recorded trace in an event
% log, where only the events are known.  Runs alternate between state
% and events, while traces only record events.}
\rnew{
  \begin{example}[Transition System]\label{ex:transition_system}
    In the transition system shown in \cref{fig:ts}, the initial state
    is indicated by an arrow with an empty source, and the final
    states are double-lined. A run in this transition
    system could be
    $r = \langle \mathsf{start}, \mathit{answer\ Q1}, \mathsf{Q1\ answered},
    \mathit{answer\ Q2},$ $ \mathsf{Q2\ answered}, \ldots \rangle$, with the corresponding
    trace
    $\pi(r) = \langle \mathit{answer\ Q1}, \mathit{answer\ Q2}, \ldots \rangle$.
  \end{example}
}

\rnew{We assume that user journeys have a sequential structure; i.e.,
  users only engage in one action at a time and logs do not contain
  noise.  With this assumption, transition systems suffice as 
  underlying formalism for user journeys.}

\subsection{Process Mining}
Process mining (PM)~\cite{DBLP:books/sp/Aalst16}
% combines formal and data science techniques support tools a
% toolchain
supports the model-based analysis of
% (business)
processes.
% . One part of the toolchain is
In PM, \emph{process discovery (PD)} techniques extract models of
processes, e.g., Petri nets, from a
% given
sample of process observations, e.g., event logs.
\citet{kobialka2025decision,kobialka24sosym,kobialka24abs} recently applied
PD techniques to  automatically generate behavioral models of user
journeys from event logs,
% extract user journey models from event logs,
based on
the \emph{directly-follows graph} (DFG) construction of
\citet{DBLP:conf/centeris/Aalst19}. 
\rnew{Well-established process discovery algorithms include, e.g.,  approaches based on genetic algorithms \cite{de2007genetic},  \emph{Inductive Miner}~\cite{leemans2013discovering} for learning process trees,  \emph{Alpha Miner}~\cite{van2004workflow} for learning workflow nets, and  \emph{Heuristics Miner}~\cite{weijters2006process} for dealing with noise.
We focus on  DFGs as they represent a fundamental modelling technique in process discovery.
%We assume that 
DFGs are well-suited for user journeys as they are sequential and 
%require 
their state-based representations can be used for,
% automated user-centric analysis techniques
e.g., model checking~\cite{kobialkaFM24,kobialka2025decision}.}
\looseness=-1

We review PD techniques within PM to generate DFGs for user journeys.
In short, PD consists of three phases: (1)~\emph{preprocessing}, which
prepares the event log; (2)~\emph{construction}, which generates the
behavioral model; and (3)~\emph{postprocessing}, which prepares this
% the generated
model for further analysis.

\textbf{Preprocessing.}
The event log~$L$ is modified according to a target granularity.
Typically, $L$ is filtered for outliers (i.e., rare journey
variations) by removing spurious traces.  Rare events may be filtered
out, but this can remove too many user interactions at this early
stage of the discovery process~\cite{DBLP:conf/centeris/Aalst19}.
Note that a preprocessed $L$ is still an event log.

\textbf{Construction.}
Commonly, a DFG is a TS $G = (\Gamma, A, E, \{s_0\}, T)$ where
$\Gamma = A \cup \{s_0\}$, transitions connect consecutive events, so
two states $a,b \in \Gamma$ are connected, $(a, b, b) \in E$, if the
log contains a trace where $a \in A$ is directly followed by
$b \in A$. The state $s_0$ is a designated initial state with outgoing
transitions to the first event of the traces in the log, and $T$
contains the observed last states in traces (or designated final
states introduced in the preprocessing).  For process analysis, the
DFG can be used to highlight the most common
behavior~\cite{DBLP:conf/centeris/Aalst19}.  \emph{Directly follow
  systems} (DFS) introduce richer state representations than DFGs by
considering additional context and different context representations
\cite{van2010process}.  DFSs have been used to model the interactions
between service providers and
users~\cite{kobialkaWeightedGames,kobialka2025decision,kobialka24sosym}.

\textbf{Postprocessing.} 
This phase consists of \emph{filtering} rare transitions and
\emph{completing} the constructed DFS. The filtering of rare
transitions can be done in terms of their frequency in the log,
adjusting the neighboring transitions traversed by the traces
iteratively if needed to preserve model
soundness~\cite{DBLP:conf/centeris/Aalst19,leemans2019directlyTool}.
Completing a DFS introduces common process patterns that are not found
in the log; e.g., enabling event interleavings or alternative
executions can be done by closing ``diamond'' patterns and resolving
self-loops~\cite{van2010process}.

\subsection{Automata Learning} 
% AL aims to learn a minimal behavioral model from a finite set
% of system traces. AL techniques may be active or passive, depending on
% how the set of traces is generated.  To learn behavioral models of
% user journeys from event logs, we use passive learning to generate a
% minimal behavioral model.\looseness=-1
AL aims to learn a minimal automaton from a finite set of system traces. \rnew{AL stems from the problem of learning an unknown regular language from a set of traces.} AL can be active or passive, depending on the set of traces that are actively generated by interacting with the system under learning (SUL) during learning or (passively) given traces.
% To learn behavioral models of
For user journeys, \rnew{we cannot generally assume access to a SUL. Hence, we discard active AL
  % to generate the user journeys
  and} use passive learning to generate a behavioral model from event logs. \rnew{Gold~\cite{DBLP:journals/iandc/Gold78} showed that learning a regular language requires positive and negative traces, where positive traces are those  accepted by the language and negative traces are those that are not accepted. However, Angluin~\cite{Angluin88} showed that the underlying stochastic distribution of events can be used to learn models from only positive traces. Since event logs for user journeys only contain positive traces, we learn probabilistic models.}
% we approach the NP-complete problem of learning a minimal behavioral
% model from a given set of traces, i.e., passive learning.
\rnew{Alergia~\cite{DBLP:conf/icgi/CarrascoO94} is a state-merging algorithm
  for learning behavioral models of probabilistic systems from
  % a set of
  positive traces, where}
state-merging is a passive AL technique for learning behavioral models of black-box systems from a set of traces.

Let $\mathcal{L}({\mathrm{SUL}}) \subseteq A^*$ be the language
defined by the SUL. Given a set of traces
$L \subseteq \mathcal{L}({\mathrm{SUL}})$, an AL algorithm constructs a
\emph{frequency prefix tree acceptor} (FPTA) $\mathcal{T}$, where
states in the tree represent events in the traces of~$L$. Let
$\sigma \ll \sigma'$ denote the reflexive prefix relation on traces,
expressing that $\sigma$ is a prefix of $\sigma'$. The FPTA
$\mathcal{T}$ is constructed such that $\sigma \ll \sigma'$ holds for
every pair of traces, where $\sigma \in A^*$ is a prefix generated from the root of
$\mathcal{T}$ and $\sigma' \in L$.  Transitions in $\mathcal{T}$
% the FPTA
are labeled
% denoting
by the number of traces from $L$
% that share
with the same prefix in $\mathcal{T}$.
% the FPTA.

AL
% Automata learning then
merges states of similar events in
% the FPTA 
$\mathcal{T}$
% under the assumption of 
by assuming the Markov property; i.e.,
% the sequence of
subsequent events depend only on the current event. 
State merges are based on an \emph{evaluation} function that allows to assess whether states can be merged, but also to create a ranking of possible merge candidates.

Alergia~\cite{DBLP:conf/icgi/CarrascoO94} is a well-established passive AL
% automata learning
algorithm based on state-merging.
% of two states in $\mathcal{T}$
% the FPTA
Alergia's evaluation function for merging states
uses the Hoeffding
bound~\cite{Hoeffding1994}:
% according to which
two states $s,s'$ in $\mathcal{T}$
% are different
differ if, for any event $a \in A$,
 \begin{equation}\label{eq:hoeffding}
  \left| \frac{f^s_a}{n_s} - \frac{f^{s'}_a}{n_{s'}} \right| > \sqrt{\frac{1}{2} \log \frac{2}{\alpha}} (\frac{1}{\sqrt{n_s}} + \frac{1}{\sqrt{n_{s'}}})
\end{equation}
holds, where $n_s \in \mathbb{N}$
% defines
is the sum of the frequencies on the outgoing transitions from state
$s$ and $f^s_a$
% corresponds to
the frequency of the current event $a$ in $s$. The
parameter $\alpha \in (0, 1]$ expresses the confidence that the
% underlying
distribution of events in the log reflects the system's behavior.  An
event log is \emph{well-distributed} if the events in the log are
statistically significant, i.e., the traces in the log follow the
underlying unknown distribution of events of the SUL.  High values for
$\alpha$ express low confidence in the distribution of the event log
% in the distribution of events in the event log
and thus favor little merging in the FPTA, while low values
% for $\alpha$
favor eager merging.  After state merging, the transition frequencies
are converted to probabilities, reflecting
% corresponding to
the proportions of the sum of frequencies on the outgoing
transitions. The
% resulting model
result is a Markov
chain~\cite{DBLP:books/daglib/0095301} 
% following a similar in structure to the Markov chain in
(see \cref{fig:running-example}).  To obtain a TS \rnew{from the learned Markov chain}, we then erase the
probabilities on the transitions and extend the transition relation by
the event of the target state. In the generated TS, all states are
considered final.\looseness=-1

\section{Mining Behavioral Models of User Journeys}\label{sec:ALvsPM}

This section demonstrates by  example how behavioral
models for user journeys can be learned using methods from AL or
PM. We also show how, depending on their parameter settings, both
methods can generate the underlying model.

\begin{figure*}[!t]
  \centering
  \begin{subfigure}[b]{0.24\linewidth}
    \centering
    \includegraphics[trim=4 5 5 75, clip, width=\linewidth]{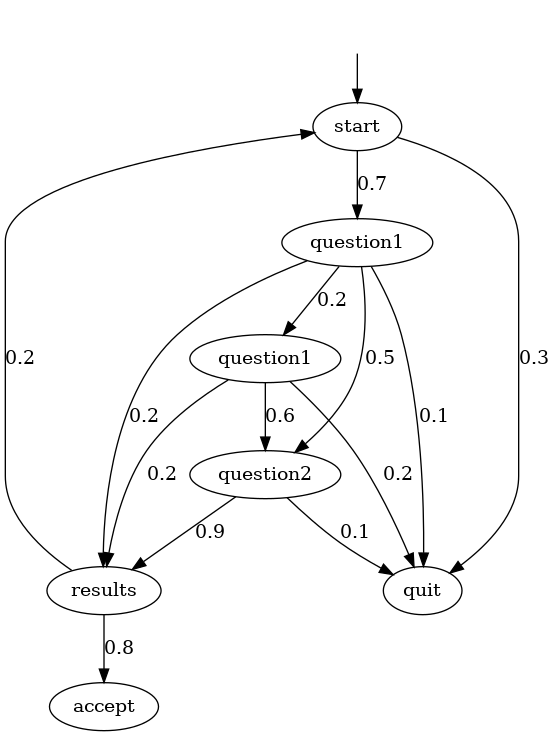}
    \caption{Ground-truth Markov chain of the assessment system (\cref{ex:survey_intro}).}
    \label{fig:running-example}
  \end{subfigure}
  \hfill
  \begin{subfigure}[b]{0.25\linewidth}
    \centering
    \includegraphics[trim= 0 0 0 50pt,clip, width=0.78\linewidth]{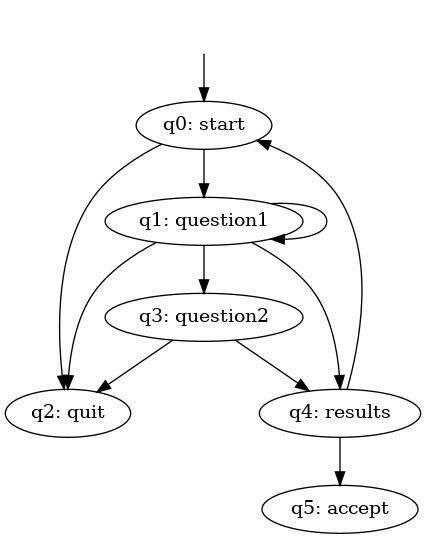}
    \vspace{10pt}
    \caption{TS from both high-confidence AL ($\alpha$\,=\,$0.1$) and PM-generated DFG\,($S^{[t]_1}$).}
    \label{fig:large-high}
    \label{fig:dfg}
  \end{subfigure}
  \hfill
     \begin{subfigure}[b]{0.18\linewidth}
       \centering
       \includegraphics[trim= 0 0 0 60pt,clip,width=1.15\linewidth]{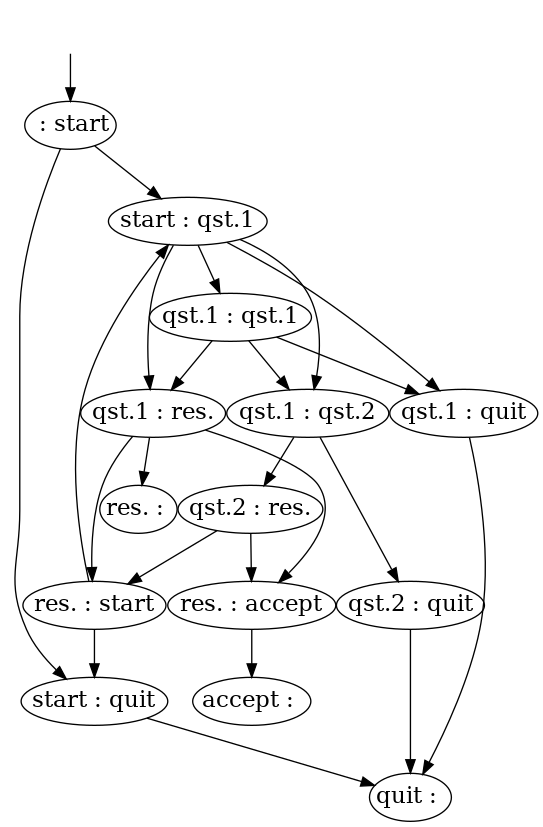}
       \caption{PM-generated DFG with 1-event trace suffix
         ($S^{[t]_1}_{[h]_1}$).}
       \label{fig:2sequence_future_success}
     \end{subfigure}
     \hfill
    \begin{subfigure}[b]{0.22\linewidth}
      \hspace{-5pt}\includegraphics[trim= 4 0 10 60pt,clip,width=1.05\linewidth]{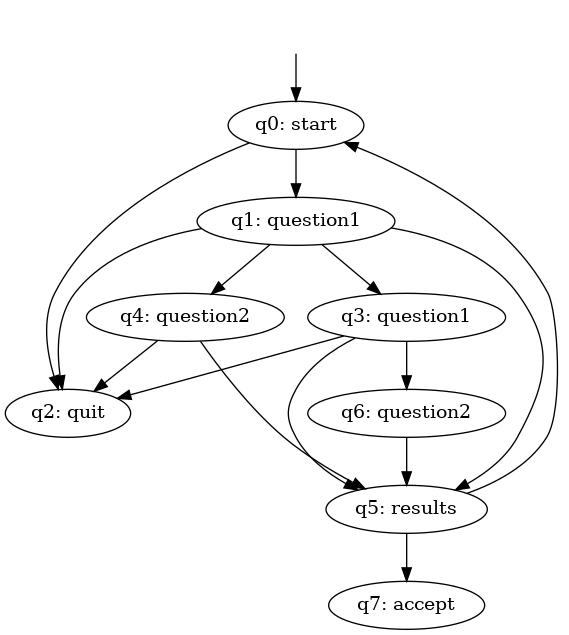}
      \vspace{5pt}
      \caption{TS from low-confidence AL
        ($\alpha = 0.9$).\\[2pt]\quad}
        \label{fig:large-low}
    \end{subfigure}
    \caption{Markov chain, learned PM and AL models of an
      assessment system. Different state representations (for PM) and
      confidence values (for AL) are used to learn models from a log with 80 traces. \rnew{For 
      %readability, 
        simplicity, we omit transition labels.}}
   \label{fig:PM_eg}
   \label{fig:alergia-website-models}
\end{figure*}

\begin{example}[Assessment System]
  \label{ex:survey_intro}
  GrepS is a company that offers programming skill assessments as a
  service. The generation of user journey models from the event logs
  of GrepS has previously been studied by \Citet{kobialkaWeightedGames}. This example presents a
  simplified version of a user journey model, based on the GrepS
  service. \Cref{fig:running-example} shows a Markov chain of user journeys, where users are asked to answer questions to obtain a
  programming skill evaluation. A user journey always begins with a
  start event.  After attempting to answering Question~1, the user can
  improve their answer once by repeating Question~1, or move to
  Question~2. Question~2 cannot be repeated.  Users may quit the
  evaluation process or request their results at any point after
  starting the evaluation. After receiving the results, the user
  accepts them or repeats the whole questionnaire.  A user journey is
  considered successful if the user accepts the evaluation results and
  unsuccessful otherwise. The Markov chain includes probabilities for
  subsequent events. \looseness-1
\end{example}

\Cref{ex:survey_intro} already illustrates the challenge of automatically
generating user journey models, since the example includes looping
behavior and repeating events.  Furthermore, the example also shows that
user behavior may differ.

\subsection{Event Logs for User
  Journeys} \label{sec:event-logs-for-user-journeys}
  
User journeys are goal-orientated processes, where users interact with
a service provider to reach a certain goal, e.g., receiving a
programming skill evaluation.  Event logs for user journeys record all
interactions with different users, i.e., both successful (when the
goal is reached) and unsuccessful user journeys.  Therefore, a trace
in an event log is a sequence of observable events that reflect a
user's interaction with the service provider or actions along the user
journey.

\begin{example}[Event Log for the Assessment System]
  A possible event log for user journeys derived from
  \cref{ex:survey_intro}:
  \begin{align*}
    &\langle \textit{start}, \textit{question1}, \textit{question2},  \textit{results}, \textit{accept} \rangle & \text{(UJ-01)}\\
    &\langle \textit{start}, \textit{question1}, \textit{results}, \textit{accept} \rangle  & \text{(UJ-02)}\\
    &\langle \textit{start}, \textit{question1}, \textit{question1}, \textit{quit} \rangle & \text{(UJ-03)}
  \end{align*}
  The traces (UJ-01) and (UJ-02) are successful user journeys,
  indicated by $\textit{accept}$ as their last event. (UJ-03) is an
  unsuccessful journey, where the user quits early.
\end{example}
  
\subsection{Process Mining for User
  Journeys} \label{sec:pm-for-user-journeys}

We extend the DFG construction from a preprocessed event log $L$
(\cref{sec:preliminaries}, Phase 2) to \emph{directly follows systems}
(DFSs) by means of two functions: \emph{trace projection} and
\emph{structure
  projection}~\cite{van2010process,DBLP:books/sp/Aalst16}.  The
purpose of the trace projection is to create a trace abstraction,
while the purpose of the structure projection is to map the abstracted
trace to states of a TS. Together, the two functions define the
\emph{state representations}.
  
\emph{Trace projections} commonly (1) filter the horizon of the trace,
e.g., a sliding window abstraction and (2) filter events, i.e., decide
to ignore certain events.  \emph{Structure projections} define state
representations from (abstracted) traces.  Structure projections
typically map traces to \emph{lists}, \emph{sets}, or
\emph{multisets}, thereby deciding how the underlying states are
generalized: Lists reflect the traces as observed, multisets abstract
from the order of the events, and sets abstract from their frequency.
Formally, the monoid $(M, \otimes)$ over events $A \subseteq M$ with
neutral element $0$ defines a \emph{structure projection}
$\oplus : A^* \to M$ for trace
$\sigma = \langle a_0, \dots, a_{n} \rangle$, with
$\oplus(\epsilon) = 0$, and
$\oplus(\sigma) = \otimes(\oplus(\sigma_{0:n}), a_{n})$.  \emph{State
  representations}~\cite{van2010process} define states in a TS from
(abstracted) traces by composing the trace and structure
projections.\looseness=-1
  
  \begin{example}[State representations]
    \label{example:state_representation}
    Let $\head_k : A^* \to A^*$ be a trace
    projection that maps a trace
    $\sigma = \langle a_0, \dots, a_n \rangle$ to its first $k$ events
    $\sigma_{:k}$, ignoring subsequent events (and assuming
    $k\leq n$).  Similarly, let $\tail_k : A^* \to A^*$ map $\sigma$
    to the tail $\sigma_{\max(0,n-k):}$, ignoring (up to) the first
    $k$ events. We define structure projection $\oplus$ on the monoid
    $(M, \otimes)$ of lists, where $A \subseteq M$ and $\otimes$ is
    list concatenation.  A trace
    $\sigma = \langle a, b, a \rangle \in A^*$ is mapped by
    $\oplus(\sigma)$ to the list $[a, b, a]$ by concatenating single
    events.
    The state representation ${[t]}_{k}$, which composes the trace
    projection $\tail_k$ with $\oplus$, only considers the list of the
    last $k$ events, i.e., states with the same last $k$ events in
    sequence are considered equal.  The state representation
    ${[h]}_{k}$, which composes the trace projection $\head_k$ with
    $\oplus$, equates traces if the next $k$ events are the same,
    resulting in a $k$-step look-ahead state representation.  By
    mapping to the monoid of sets, we obtain the state representation
    ${\{t\}}_{k}$ which ignores the position and frequency of the last
    three elements, e.g.  ${\{t\}}_{3}(\sigma) = \{a, b \}$.
\end{example}

A DFS can be efficiently constructed by applying the DFG construction
(see \cref{sec:preliminaries}) to state representations.  While states
in a DFG only depend on the previous event in the trace prefix, states
in the DFS depend on state representations for the trace prefix and
suffix over $k$ events, respectively. Thus, the state representation
becomes sensitive to both past and future events (see
\cref{example:state_representation}). Depending on the state
representation for the trace suffix, the DFS might contain several
initial states.  \looseness=-1

  \begin{definition}[DFS from state representations]
    \label{def:dfs_def}
    Let $L$ be an event log over events $A$ and let $r$ and $r'$ be
    state representations.  A \emph{directly follows system} (DFS)
    over $L$ that uses $r$ as a prefix and $r'$ as a suffix
    representation, is a transition system
    $S^{r}_{r'} = (\Gamma, A, E, I, T)$ such that
    \begin{itemize}
        \item $\Gamma = \{ (r(\sigma_{:k}), r'(\sigma_{k:})) \mid
        \sigma \in L, 0 \leq k \leq \lvert \sigma \rvert \}$, 
         \item $A = \{ a \mid (s, a, s') \in E \}$,
        \item $E = \{ \langle (r(\sigma_{:k}), r'(\sigma_{k:})),
        \sigma_{k+1}, (r(\sigma_{:k+1}), r'(\sigma_{k+1:})) \rangle \mid \sigma \in L, 0 \leq k \leq \lvert \sigma \rvert \}$, 
          \item $I = \{(r(\epsilon), r'(\sigma)) \mid \sigma \in L \}$, and
          \item  $T = \{ (r(\sigma), r'(\epsilon)) \mid \sigma \in L \}$.
    \end{itemize}
  \end{definition}
  
  \rnew{DFSs can be constructed similarly to DFGs after constructing the states, by applying the state representation functions to every event $\sigma_i \in \sigma$ for all traces in a log $L$.
  While the state computation must be applied to all $\sigma_i$, it can be efficiently computed as the horizon used in the trace projections is usually small, e.g., a finite number of previous events.
  % To compute a DFS, we apply the state representation to each $\sigma_i$ in  $\sigma$ for all traces $\sigma \in L$.
  Further improvements like filtering for unique traces and efficient log representations can further speed-up this computation.}
  
  \begin{example}[PM for the Assessment System] \label{ex:pm-for-uj}
    We consider different state representations for an event log with
    80 traces, generated by sampling the Markov chain from Example~1,
    where an event is sampled with the probability of its associated
    transition.  Thus, traces in the log have different probabilities,
    to reflect event logs of user journeys. We select the length of
    the traces uniformly at random, between two and 15 events.
    Independent of the selected maximum length, a user journey stops
    after an \emph{accept} or \emph{quit} event. Every trace starts
    with an initial \emph{start} event.  \Cref{fig:PM_eg} presents
    DFSs for different state representations for the survey system:
    \cref{fig:dfg} shows the DFG, only considering the most recent
    event (technically, the state representation $S^{[t]_1}$). \rnew{Note that we omit transition labels, but they can be obtained from the state labels as shown in \cref{fig:ts}.}
    \Cref{fig:2sequence_future_success} presents the DFS considering the
    current event and the next event (i.e., $S^{[t]_1}_{[h]_1}$).
    Observe that the DFG ($S^{[t]_1}$) does not recognize that the
    initial question can only be answered once, but displays it as a
    self-loop; i.e., users cannot improve their answers arbitrarily
    often. In contrast, the DFS $S^{[t]_1}_{[h]_1}$ covers the
    original model and displays that the answer to question~1 can only
    be improved once, thereby covering the different possibilities to
    answer the two questions.\looseness=-1
  \end{example}

  DFSs satisfy the following properties: (1) every DFG is a DFS, (2)
  there are no dead transitions or states, i.e., every state can reach
  a final state, and (3) all log traces are traces of the generated
  DFS.  \looseness=-1

  \subsection{Automata Learning for User
    Journeys} \label{sec:automata-learning-for-user-journeys} We
  generate behavioral models of user journeys from event logs using
  passive AL. Classic
  passive learning techniques require a log with traces that are part of the language of the system under learning as well as traces that are not part of the language.~\cite{DBLP:journals/iandc/Gold67}. \citet{agostinelli2023process}
  compared different AL algorithms for mining \emph{deterministic
    finite automata} of business processes. Their work separates the event log into traces that are in the language and traces that are not, according to trace length. For learning behavioral models of user journeys, this
  approach is less suitable since later model analysis might be
  interested in investigating the length of user
  interactions. Classifying according to the success of the user
  journey could also limit the insights gained, as user journey models
  should also allow conclusions to be drawn about the reasons for not
  achieving a goal. Thus, we want to learn a behavioral model whose
  language includes all traces of a given event log.

  \citet{Angluin88} observed that passive learning from only positive
  examples is possible when considering the probability distribution
  of the underlying event log. AL algorithms of this kind
  % Alergia~\cite{DBLP:conf/icgi/CarrascoO94} 
  %\red{is an example of an AL algorithm that} 
  %require only positive
  % examples,  where the 
  generate a Markov chain, which can then be translated
  to a TS \rnew{by erasing the transition probabilities}, as described in \cref{sec:preliminaries}. Alergia~\cite{DBLP:conf/icgi/CarrascoO94} is an example of a
  state-merging algorithm that evaluates possible state-merges according to the Hoeffding-bound. For this, Alergia assumes
  that the provided log follows a certain probability distribution. However, this introduces the challenge of determining whether a given event log contains
  enough user journeys to distinguish states when applying AL for
  learning user journey models.

  The Hoeffding-bound (\cref{eq:hoeffding}) considers $\alpha$ as a
  parameter to steer our confidence in the event log following a
  certain distribution. The parameter $\alpha$ is an arbitrary small
  real within $(0,1]$ defining the error, such that a higher $\alpha$
  expresses less confidence in the distribution of the underlying
  event log. For learning user journeys, we prefer a cautious state
  merging, i.e. a higher $\alpha$, based on the following assumptions
  about user journeys. First, we assume that user behavior implements
  a regular language. The regular language defining a user journey
  includes in contrast to classical language identification problems,
  e.g., Tomita grammars~\cite{tomita:cogsci82}, at least one order of
  magnitude more possible events in the considered alphabet. Second,
  user journeys frequently follow a strict sequential structure of
  events towards achieving the desired goal. This might be explained
  by the fact that the degree of freedom of actions taken by the user
  to interact with a service provider is usually limited and that
  following events are only useful for achieving a goal after a
  certain sequence of previous events.
  \looseness=-1
  
\begin{example}[AL for the Assessment System]\label{ex:al-for-uj}
  \Cref{fig:alergia-website-models} depicts models learned by Alergia
  with different values for $\alpha$.  The event log is generated from
  the assessment system of Example~1.  We evaluated a higher
  confidence ($\alpha = 0.1$), \cref{fig:large-high}, and a lower
  confidence ($\alpha = 0.9$), \cref{fig:large-low}, on the log with
  80 traces from \cref{ex:pm-for-uj}.  Alergia generates Markov
  chains, which we translate into TSs.  The results show that learning
  with $\alpha\!=\!0.1$ (\cref{fig:large-high}) generates an
  overapproximation, i.e., the model describes a more general language
  than the ground truth.  The reason is that low $\alpha$ promotes
  state merging since more states are similar according to
  \cref{eq:hoeffding}. Note that this model is the same as the DFG
  generated in \cref{ex:pm-for-uj}, which does not distinguish the two
  Question~1 events. By an increase to $\alpha = 0.9$, Alergia merges
  fewer states.  The corresponding model (\cref{fig:large-low}) is an
  underapproximation; i.e., the model cannot produce all traces
  observable in the ground truth.
\end{example}
  
\Cref{ex:al-for-uj} shows that different experimental setups influence
the resulting models. Thus, \cref{ex:pm-for-uj,ex:al-for-uj}
demonstrate the impact of the event log on both PM and AL. For PM,
creating a DFG might lead to an overapproximation; however, selecting
the right size for trace prefix and suffix requires expert domain
knowledge. Therefore, PM techniques usually apply preprocessing to
reduce and adapt the log. For AL, a correct model requires the right
assumption about the underlying distribution of events. Otherwise, AL
may learn an over- or underapproximation.  The degree of generality of
the model depends strongly on the underlying event log, and poses a
challenge for finding appropriate parameter settings for the learning
algorithm.  We address these challenges in \cref{sec:hybrid}.
\looseness=-1

\section{A Hybrid Learning Method Combining PM and AL} 
\label{sec:hybrid} 
We introduce a novel learning method, called
\name{Hybrid}, that combines AL and PM techniques to overcome the
limitations of these techniques with respect to the quality of the
underlying event log. This technique automatically selects the best
suited method, depending on the characteristics of the log.  
\name{Hybrid} is a general method independent of specific PM and AL methods. It can be instantiated with any PM and AL methods that allows to infer a TS from an event log.
In the following, we present \name{Hybrid} based on the PM and AL methods introduced in \cref{sec:ALvsPM}.
By design, the PM method (\cref{sec:pm-for-user-journeys})
risks to create an underapproximation on large logs, as the selected
state presentation might not enable a sufficient generalization of the
behavior. In contrast, AL (\cref{sec:automata-learning-for-user-journeys}) tends to learn an
overapproximation, as AL techniques are designed to create the minimal
automaton by merging as many states as possible. This might lead to
overly general models, due to the sparsity of the provided event
log. \name{Hybrid} will be validated by the experiments in
\cref{sec:syn-exp,sec:real-exp}.\looseness=-1

\name{Hybrid} applies PM on small
% event logs
and AL on well-distributed event logs. The idea is to switch from PM
to AL if the
% underlying
event log is sufficiently well distributed. For AL, state merging is
guided by the confidence parameter $1 - \alpha$. Therefore,
\name{Hybrid} aims to automatically switch from PM to AL if confidence
in the event log being well distributed is high enough.  This decision
is implemented in a \emph{certainty threshold} $\lambda$ such that the
% behavioral
model is constructed using PM for $\lambda>\alpha$,
otherwise AL is used.\looseness=-1

To estimate the certainty threshold $\lambda$, we propose a function
$\lambda_\mathrm{approx}$ based on three different aspects of an event log
$L$ that influence our confidence $\alpha$: (1) the number of different
traces in an event log $L$~\rnew{\cite{van2022foundations} (i.e., the number of trace variants recorded in $L$: $\var(L) = |\{\sigma \in A^* \mid \sigma \in L \}|$, where $\var(L)$ is logarithmically scaled}) and (2) the event log size
$\lvert L \rvert$. We further introduce (3) a coefficient $C_0$ to adapt
the function in the presence of domain knowledge. Let
$\lambda_\mathrm{approx}$ be defined as follows:\looseness=-1
\begin{equation}\label{eq:hybrid}
  \lambda_\mathrm{approx} = \frac{C_0 \cdot \log_{10}(\var(L))}{\lvert L \rvert}.
\end{equation}

For user journeys,
we set $C_0 = \lvert A \rvert$, \rnew{accounting for the number of distinct events}.
Observe that for logs with a fixed event set size $\lvert A \rvert$,
$\lambda_\mathrm{approx}$ shrinks rapidly in the size of the log,
thus, returning a clear decision point between AL and PM. The
intuition behind \cref{eq:hybrid} is as follows: Small log sizes
result in low confidence that the log is adequately distributed,
especially for large event set sizes. In this case, we target the usage
of PM techniques. However, we want to switch to AL as early as
possible to avoid underapproximations caused by PM. By letting
$\lambda_\mathrm{approx}$ shrink rapidly and $\alpha$ decrease with
increasing log sizes, \name{Hybrid} is able to learn behavioral models
of user journeys to avoid overapproximations as explained in
\cref{sec:automata-learning-for-user-journeys}.

When selecting $\alpha$ values for learning user journey models with
AL, our goal is to find parameters for the AL setup that avoid both
overapproximation and underapproximation in the learned
models. Investigating the impact of different $\alpha$ values on AL,
\citet{DBLP:journals/ml/MaoCJNLN16} show that with a sufficiently
large event log $L$, $\alpha = \frac{1}{N}$ can be a good
approximation, where $N = \sum_{\sigma \in L} \lvert \sigma
\rvert$. However, we cannot generally assume that event logs for user
journeys are large enough to use this approximation. Therefore, we
propose an alternative $\alpha$ setup to what is normally done for AL,
and adopt a sigmoid function \looseness=-1

\begin{equation}\label{eq:alpha}
  \alpha_\mathrm{approx} =
  1 - \frac{1}{1 + e^{(-\lvert L \rvert + C_1)\, C_2}}.
\end{equation}

\noindent
In this function, the constant $C_1$ shifts the inflection point, and
the coefficient $C_2$ adapts the slope. By choosing $C_1$ and $C_2$
appropriately, the function allows an $\alpha$-selection based on the
size of $L$. The $\alpha$-values that are close to $1$ for very small
logs $L$ and approach $\frac{1}{N}$ for larger $L$.

\begin{example}[$\alpha$-Approximation for AL]\label{ex:alpha-for-al}

  For the event log with 80 traces in \cref{ex:pm-for-uj},
  \cref{eq:alpha} calculates $\alpha_\mathrm{approx} = 0.49$ by
  setting the coefficients according to the event set size
  $\lvert A \rvert = 6$, where $C_1 = {10 \cdot \lvert A \rvert}$ and
  $C_2 = \frac{1}{100 \cdot \lvert A \rvert}$. When using
  $\alpha_\mathrm{approx} = 0.49$ for $\alpha$, AL 
  learns a TS which is equivalent to the TS representation of the
  Markov chain shown in \cref{fig:running-example}. Note that with the
  standard approximation $\alpha=\frac{1}{N}$ and this event log, the
  resulting value for $\alpha$ would be lower than $0.1$, which would lead to an overapproximation similar to the one shown in \cref{fig:large-high}.  \looseness=-1
  
\end{example}

%%% Local Variables:
%%% mode: latex
%%% TeX-master: "main"
%%% End:

\section{Experiments on Synthesized Benchmarks} \label{sec:syn-exp}

% \red{
% \highlight{RQ1:}{How well do the behavioral models learned by AL and PM techniques represent ground-truth models?}
% }
% \noindent\begin{minipage}{0.98\textwidth}\fbox{\parbox{\textwidth}{
% \textbf{RQ1:} How well do the behavioral models learned by AL
%   and PM techniques represent ground-truth models?
% }}\end{minipage}}

In this section, we evaluate our hybrid method and compare it to
different AL and PM configurations. For this purpose, we develop a
novel benchmark suite for user journeys, which allows comparisons with
``ground-truth'' models. To this aim, the benchmark suite includes
synthesized behavioral models of user journeys, \rnew{represented as process trees}, and the corresponding event logs \rnew{sampled from the process trees}. For AL, we evaluate different setups for the parameter
$\alpha$. For PM, we investigate different abstractions and state
representations.  By comparing with the ground-truth models, we can
evaluate the precision and recall of the user journey models generated
by AL and PM and assess the degree of over- or under-approximation. In
this section, we answer the following research question:

\medskip

\noindent
\emph{\textbf{RQ1:} How well do behavioral models learned by AL and PM techniques represent ground-truth models?}

\subsection{Experimental Design and Setup}\label{sec:machine}
We implemented our evaluation framework in Jupyter Notebooks, using Python 3.10.12.
The generated benchmark set and the code is available
online.\footnote{\url{https://figshare.com/s/3acae9725dda09cce811}} Due to the numerous
experiments, we distributed the executions over (1) a laptop with $32\unit{GB}$ memory and an i7-1165G7 @
$2.8\unit{GHz}$ Intel processor, and (2) a workstation with
$128\unit{GB}$ memory and a 64-core AMD EPYC processor.\looseness-1 % For AL, we used the Python library
%\textsc{AALpy}~\cite{DBLP:journals/isse/MuskardinAPPT22}, version
%1.4.1. \looseness-1

\textbf{Synthesizing User Journeys.}
We aim to synthesize behavioral models of user journeys, therefore,
our generation incorporates structural elements that can be found in
real-world user journeys. The PM library
\textsc{PM4Py}~\cite{bertiPM4PyProcessMining2023} enables the
randomized creation of models such as Petri nets. \textsc{PM4Py}
implements also a generator for process trees as proposed by
\citet{jouckPTandLogGenerator}. 
Process trees are inherently sound, can be generated stepwise, and can be restricted to the same languages as transition systems~\cite{jouck2019generating}.
The process tree generator allows the regulation of the creation of behavioral characteristics via parameters.
For the creation of user journeys, we consider the following four control-flow
features: (1) sequential structures, (2) branch statements, (3) loops,
and (4) duplicate event symbols. We either enable or disable these
features. Considering all parameter combinations, we generate $14$
different setups (excluding the setup where all features are disabled
and only duplicate events are allowed). For all generated process
trees, we disable certain behavioral concepts that do not apply to
user journeys, e.g., parallelism and silent transitions since user
behavior is assumed to be sequential and observable. Furthermore, we
align the size of the event alphabet to the real-world examples
presented in \cref{sec:real-exp}, where the number of
different event symbols ranges between $20$ and $30$. From each
generation setup, we use \textsc{PM4Py} to randomly generate $10$
process trees, leading to $140$ generated trees.

\textbf{Synthesizing Event Logs.}
\textsc{PM4Py} enables the creation of event logs from process
trees. A trace in the event log is generated by traversing through the
process tree, where the generator selects the next event uniformly at
random when several paths are enabled. To evaluate the impact of event
log sizes, we create event logs for each process tree with the
following number of traces: $10$, $50$, $100$, $200$, $500$, and
$1\,000$. For each event log size, we repeat the generation $10$
times, which sums up to $8\,400$ generated event logs in the benchmark
set.

\textbf{Process Mining Setup.} 
For PM, we implement the DFS construction from
\cref{def:dfs_def} by reducing traces to the considered length of past
(prefix) and future (suffix) events and choosing a state
representation from set, list, and multiset for pre- and suffixes
independently.  We learn DFS in three settings: (1) \name{PM-DFG}, a
DFG $S^{[t]_1}$, (2) \name{PM-long}, a four-element prefix and suffix
DFS $S^{[t]_4}_{[h]_4}$ and (3) \name{PM-uj}, a model imitating domain
knowledge by generating a DFG if it is known that no events appear
twice.  To choose one DFS $S$ among the possible configurations, we
evaluate the number of loops, $\loops_S$, and states, $|\Gamma_S|$,
and aim to minimize $\loops_S^2 + |\Gamma_S|$, thereby squaring the
number of loops to favour simpler models.  Counting loops might be
computationally expensive, so we search for DFS with
$\loops_S^2 + |\Gamma_S| < 10^6$.  If no such model is found, we
default to a DFG.
% \pknote{$S^{[t]_4}_{[h]_4}$, respectively $S^{[t]_4}$ for logs too
% large to process in (2)}

\textbf{Automata Learning Setup.}
For AL, we use the implementation of Alergia for
learning Markov chains in
% provided by
the Python
% automata learning
library \textsc{AALpy}~\cite{DBLP:journals/isse/MuskardinAPPT22}, version 1.4.3. The
learning results for Alergia depend on the parameter $\alpha$, which
regulates confidence about the underlying distribution of the log. To
evaluate the impact of $\alpha$, we learn TS with three different
$\alpha$-setups: (1) \name{AL-0.1}, higher confidence $\alpha = 0.1$,
(2) \name{AL-0.9}, lower confidence $\alpha = 0.9$, (3) \name{AL-uj},
 $\alpha = \alpha_\mathrm{approx}$ using the approximation function
 in \cref{eq:alpha}, with coefficients
according to the alphabet size $\lvert A \rvert$,
$C_1 = 10 \cdot \lvert A \rvert$ and
$C_2 = \frac{1}{100 \cdot \lvert A \rvert}$.

\textbf{Hybrid Learning Setup.}  For \name{Hybrid}, we switch between
PM and AL based on the decision threshold $\lambda_\mathrm{approx}$. We calculate
$\lambda$ according to $\lambda_\mathrm{approx}$ in \cref{eq:hybrid}
with $C_0 = \lvert A \rvert$. When PM is selected, we use PM setup (3)
\name{PM-uj}. In the case of AL, we apply the AL setup (3) \name{AL-uj}. In
both cases, we assume that these setups are the most accurate for each
method as they are designated
% \pknote{implement a targeted setup}
to generate user journey models from event logs.
\looseness=-1

\textbf{Evaluation Setup.}  The language difference between the
generated model $\mathit{TS}$ and the ground-truth model
$\mathit{TS}_\mathrm{SUL}$ is evaluated using the language comparison
technique for TSs proposed by
\citet{DBLP:journals/tosem/WalkinshawB13}. For this, we create a
confusion matrix~\cite{DBLP:journals/ipm/SokolovaL09} based on a
finite test suite $\mathcal{S} \subset A^*$. For example, we classify
a trace $\sigma \in \mathcal{S}$ as \emph{false negative (FN)} if
$\sigma \notin \mathcal{L}(\mathit{TS})$ and
$\sigma \in \mathcal{L}(\mathit{TS}_\mathrm{SUL})$. We then calculate
\emph{precision}, \emph{recall}, and the \emph{F-measure} (the
harmonic mean between precision and recall) following the standard
definitions~\cite{DBLP:journals/tosem/WalkinshawB13}. The generated
test suite provides transition coverage for $\mathit{TS}_\mathrm{SUL}$
and $\mathit{TS}$. A test sequence can be written as a triple
$(p, a, s) \in \mathcal{S}$, where $p \in A^*$ is a sequence leading
to a state, $a \in A$ is the event of currently considered transition,
and $s \in A^*$ is a randomly generated suffix of length
$[0, n_\mathrm{len})$. For each transition, we generate $n_{\sigma}$
traces. For the evaluation, we set $n_\mathrm{\sigma}$ to the
difference in the number of states between $\mathit{TS}_\mathrm{SUL}$
and $\mathit{TS}$, but at least to two. We set $n_\mathrm{len} = 2$
since the probability of generating true negative traces is high.

\begin{figure*}%
  \centering
  \begin{subfigure}[t]{.33\textwidth}\centering
  \includegraphics[trim={0 0 45 40},clip,height=40mm]{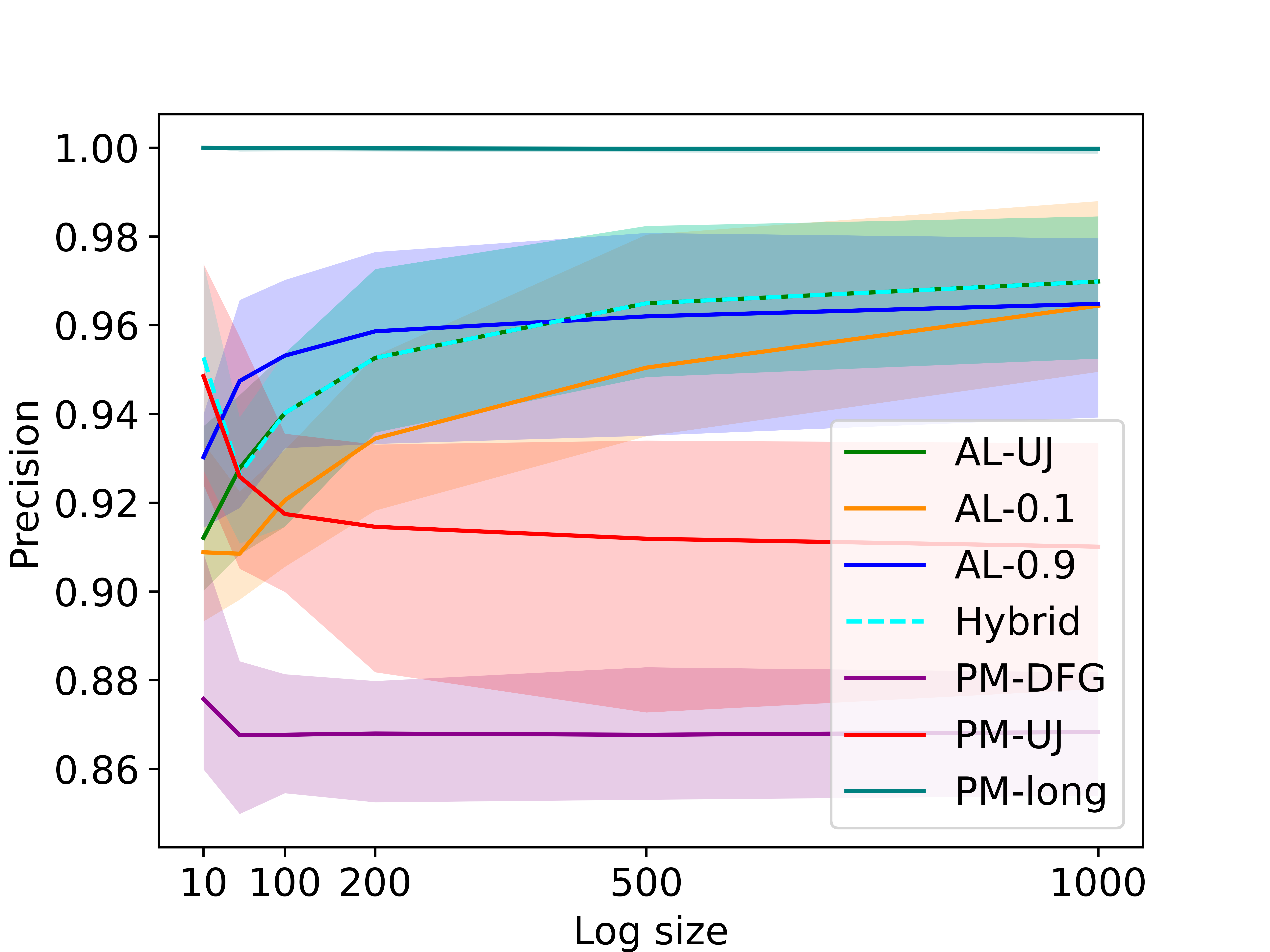}%
  \caption{Average precision\newline\label{fig:precision}}
  \end{subfigure}%
  \begin{subfigure}[t]{.33\textwidth}\centering
  \includegraphics[trim={10 0 45 40},clip,height=40mm]{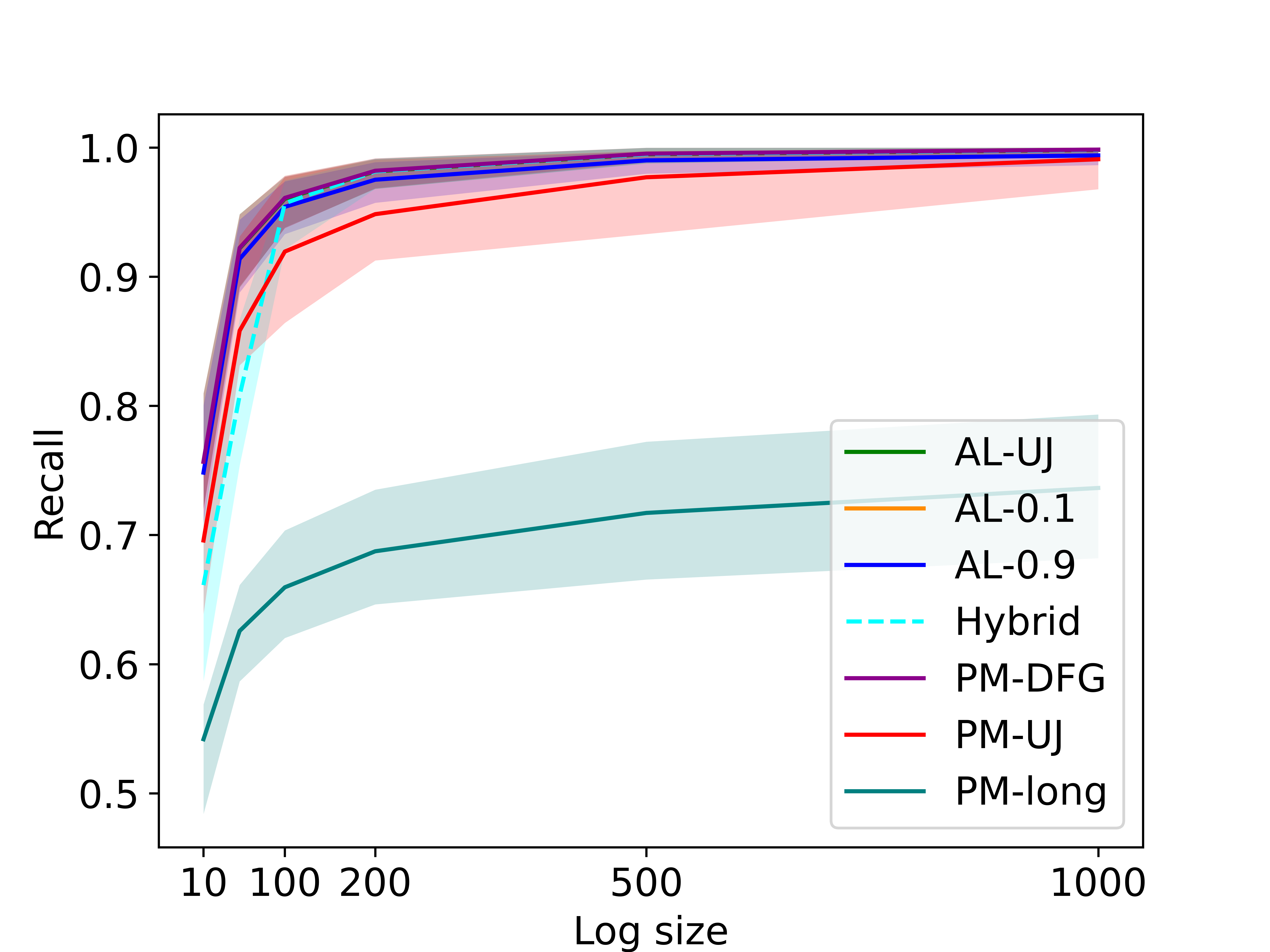}%
  \caption{Average recall\newline\label{fig:recall}}
  \end{subfigure}%
    \begin{subfigure}[t]{.33\textwidth}\centering
      \includegraphics[trim={0 0 45 40},clip,height=40mm]{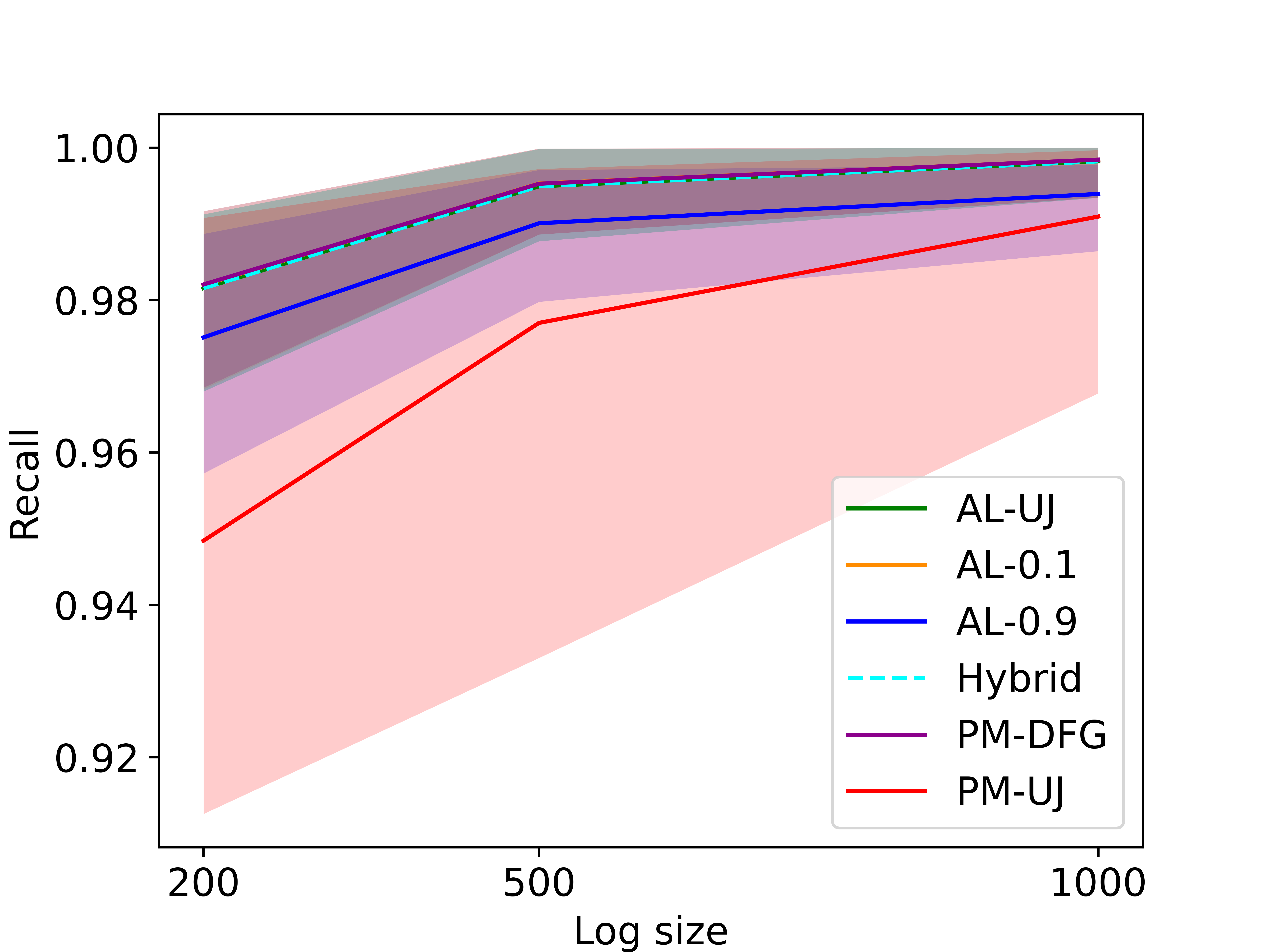}%
      \caption{Average recall (zoom $200-1000$); here, AL-0.1 and PM-DFG
        overlap\label{fig:recall-zoom}}
  \end{subfigure}
  \caption{Results for the six evaluated learning setups on the synthesized benchmark set.}
  \label{fig:pm-benchmark-results}
  \end{figure*}
 
  \subsection{Results}
\label{sec:synth-results}
\Cref{fig:pm-benchmark-results} shows the results of the learning
setups for the generated benchmark sets, for a total of $58\,800$
learned TSs.
% For better visualization,
\Cref{tab:benchmark-results} shows
% provides
the aggregated values.  The precision results for the AL experiments,
\cref{fig:precision}, show that precision increases with the size of
the
event
log, especially for \name{AL-0.1} and \name{AL-uj}. For the
recall, \cref{fig:recall,fig:recall-zoom},
% we observe that
\name{AL-0.9}
tends to underapproximate the log, since it is below \name{AL-0.1} and
\name{AL-uj}. The function $\alpha_\mathrm{approx}$
% in \cref{eq:alpha}
for \name{AL-uj} creates a good approximation for learning
models.

% \begin{wraptable}[11]{R}{0.52\textwidth}
\begin{table}
  \vspace{-14pt}
  \footnotesize
    \centering
    \caption{Average
      % benchmarks
      scores for all learning setups.}
  \label{tab:benchmark-results}
  \begin{tabular}{l@{\;}rrrrrr}
    \toprule
    & \multicolumn{2}{c}{\textbf{Precision}} & \multicolumn{2}{c}{\textbf{Recall}}  & \multicolumn{2}{c}{\textbf{F-Measure}} \\
     & avg & std & avg & std & avg & std  \\
    \midrule
    \begin{comment}
    AL-0.1 & 0.942 & 0.024 & \textbf{0.973} & 0.029 & 0.951 & 0.031 \\
    AL-0.9 & 0.962 & 0.008 & 0.966 & 0.030 & \textbf{0.960} & 0.022 \\
    AL & 0.958 & 0.017 & \textbf{0.972} & 0.029 & \textbf{0.960} & 0.026 \\
    PM-DFG & 0.865 & 0.001 & \textbf{0.973} & 0.029 & 0.901 & 0.017 \\
    PM-long & \textbf{0.999} & 0.001 & 0.703 & 0.050 & 0.777 & 0.035 \\
    PM & 0.933 & 0.009 & 0.949 & 0.054 & 0.925 & 0.033 \\
    \end{comment}
    \name{AL-0.1}  & 0.930          & 0.123 & \textbf{0.935} & 0.152 & 0.919          & 0.123 \\
    \name{AL-0.9}  & 0.952          & 0.098 & 0.929          & 0.153 & \textbf{0.929} & 0.118 \\
    \name{AL-uj}      & 0.943          & 0.108 & \textbf{0.935} & 0.152 & \textbf{0.927} & 0.119 \\
    \name{PM-DFG}  & 0.869          & 0.174 & \textbf{0.935} & 0.152 & 0.880          & 0.136 \\
    \name{PM-long} & \textbf{0.999} & 0.001 & 0.661          & 0.321 & 0.743          & 0.273 \\
    \name{PM-uj}      & 0.921          & 0.150 & 0.898          & 0.216 & 0.880          & 0.183 \\
    \name{Hybrid}  & 0.950          & 0.107 & 0.900          & 0.217 & 0.899          & 0.177 \\
    \bottomrule
    \end{tabular}
\end{table}
% \end{wraptable} 
For PM, the DFG already contains all possible transitions on small
event logs. Therefore, it does not improve with more samples. 
Compared
to the DFG, the models generated with \name{AL-0.1} overapproximate
less, since the precision of \name{AL-0.1} is higher. The results show
that considering a long prefix of events, \name{PM-long} models create
an underapproximation, i.e., the models overfit the traces in the
event log.
The \name{PM-uj} technique based on domain knowledge achieves
the best approximation for all considered PM techniques, yet tends to
learn an overapproximation compared to AL.\looseness=-1
% \pknote{PM-DFG is overapproximations, less precision. PM-long is
% clear underapproximation, perfect precisions, bad recall, PM is less
% overapproximation than than PM-long but more than AL, less recall.}

For \name{Hybrid}, we observe that selecting the setups \name{PM-uj} and
\name{AL-uj} has been the right choice, 
as both setups are best within
their learning technique. 
Furthermore, switching between PM and AL based on $\lambda_\mathrm{approx}$ provides a well-balanced solution.
In \cref{fig:precision}, we can clearly see
that \name{Hybrid} overcomes the limitations of AL and PM.
For
small log sizes, \name{Hybrid} adapts \name{PM-uj} and, therefore,
avoids overapproximating with \name{AL-uj}. However, we switch exactly at
the intersection of \name{PM-uj} and \name{AL-uj}, avoiding
underapproximations as shown in \cref{fig:recall}.

To answer \textbf{RQ1}, we conclude that AL benefits from large event
logs to learn accurate models. The sigmoid approximation
$\alpha_\mathrm{approx}$ achieves highly accurate models on
average. For PM, the state representation is crucial
% has to be carefully selected
to avoid underapproximations, preferably exploiting
% with expert
domain knowledge. By combining
\name{AL-uj} and \name{PM-uj}, \name{Hybrid}
% proves as solid method to
learns accurate models independent of the event log size.

\subsection{Threats to Validity}
\label{sec:threats_sync}
We developed our own benchmark to simulate user journeys, as we assume
that the existing benchmarks for AL and PM do not reflect the
behavioral aspects of the user journeys. The random generation of
ground-truth models still bears the risk of creating user journeys
that might be unlikely in practice. Therefore, we evaluate AL and PM
also on real-world case studies in \cref{sec:real-exp}. In addition,
although the test method covers the transitions, it does not
exhaustively test all behavior. We assume that transition coverage
sufficiently tests critical behavior of user journeys. Other testing
methods, e.g., W-Method~\cite{journals/cyb/Vasilevskii73,
  DBLP:journals/tse/Chow78} or probably approximately correct (PAC)
testing~\cite{DBLP:journals/cacm/Valiant84}, would test many negative
traces due to the restrictive sequential structure of user journeys.
% We didn't use any of the published challenge and benchmarks set
% available from AL, e.g. the \emph{Tomita} grammars, or event logs
% published in the \emph{Process Discovery Contest} (PDC)
% \footnote{https://www.tf-pm.org/competitions-awards/discovery-contest}
% to focus our evaluation on aspects unique to user journeys.  The AL
% benchmarks focus on complex grammar and PDC on complex processes
% with a highly parallel structure.
\name{Hybrid} uses a syntactic criterion to decide between PM and AL
technique, where the variance in the event log is compared to its
size.  This criterion can be extended to include the semantics of the
event log; e.g.,
% the \emph{block} or \emph{k nearest neighbor} entropy proposed by
% Back \etal
\citet{back2019entropy} propose entropies specifically for event logs.

The performance of \name{Hybrid} depends on the AL and PM methods that it uses.
We chose Alergia for the AL part as it is a well-established algorithm with a maintained implementation in publicly available learning libraries such as \textsc{AALpy}~\cite{DBLP:journals/isse/MuskardinAPPT22}.
\rnew{Alergia naturally supports the use of the confidence parameter  $\alpha$ to decide between AL and PM.}
However, in the literature on AL benchmarking~\cite{DBLP:conf/icgi/SoubkiH23}, \emph{Evidence Driven State Merging} (EDSM)~\cite{DBLP:conf/icgi/LangPP98} shows favorable performance over Alergia for classical grammatical inference problems. 

\rchange{
Recently, \textsc{AALpy}~\cite{DBLP:journals/isse/MuskardinAPPT22} has also been extended by \emph{Evidence Driven State Merging} (EDSM)~\cite{DBLP:conf/icgi/LangPP98}, which implements a state merging algorithm that is based on an evaluation function that only considers a limited amount of future states. 
}
{Recently, a general state-merging approach based on EDSM was included
  % presented and implemented
  in \textsc{AALpy}~\cite{DBLP:journals/isse/MuskardinAPPT22,von2025extending}.
A general \emph{scoring} function to rank possible merges enables the algorithm to favour merges with the largest support in the underlying log.
This implementation allows to combine evidence-drive state merging with other state-merging based algorithms, such as Alergia.
}

\rnew{In further experiments (not reported in this paper),  we tested EDSM as defined by Lang \etal~\cite{DBLP:conf/icgi/LangPP98} and EDSM in combination with Alergia, instantiated with default parameters, and it did not achieve a better performance than our reported results in \cref{sec:synth-results}. We envision that for better performance, EDSM would also require a parameter optimization, as we have done in \name{AL-uj}. Additionally, the scoring function could be adapted to include domain knowledge to further adjust state merges in user journey models.}
% In our experiments, we instantiated \name{Hybrid} also with EDSM with default parameters provided in \textsc{AALpy} as AL part and achieved a comparable performance to the reported results. For better performance, EDSM would also require a parameter optimization, as we have done in \name{AL-uj}.

%%% Local Variables:
%%% mode: latex
%%% TeX-master: "main"
%%% End:

% \input{06_practical_considerations}
\section{Experiments on Real-World Case Studies}\label{sec:real-exp}

We now evaluate the applicability of AL and PM techniques in practice.
As no ground-truth models exist, we here compare behavioral
similarities between the models learned by AL, PM, and
\name{Hybrid}.  Our results demonstrate that the quality of
learned models depends on the sparsity of the event log, and that
preprocessing may be crucial in practice.  In summary, we address how
AL and PM techniques scale to real-world user journeys by addressing
the following research questions:\looseness=-1

\medskip

\noindent
\emph{\textbf{RQ2:} What is the impact of sparse event logs on
  learning with PM and AL?\\ \textbf{RQ3:} How do the models learned
  by PM and AL for real-world user journeys compare?}

\subsection{\rnew{Real-World} Case Studies}

For the practical evaluation of how the proposed AL and PM techniques
scale to real-world case studies, we considered four different event
logs.
\rnew{The event logs are constructed from datasets by grouping events into traces and ordering the events in a trace by their timestamps.}

\textbf{GrepS.}  Event log for user journeys from the company GrepS,
for conducting programming skill evaluations with job applicants; this
case study is taken from \citet{kobialka24sosym}. The user journey
consists of (i) a sign-up phase, (ii) a task-solving phase where the
users are presented with a range of real-world programming tasks, and
(iii) a review phase, where users are presented with their evaluation
and are asked to approve their results. The provided log contains 33
traces.

\textbf{BPIC12, BPIC17a \& BPIC17b.}  BPIC, the Business Process
Intelligence Challenge, is organized regularly by the IEEE Task Force
on Process Mining.\footnote{\url{https://www.tf-pm.org/}} We use the
BPIC12~\cite{bpi2012} and BPIC17~\cite{bpi2017} event logs from a
Dutch financial institution. The logs provide detailed user
interactions, e.g.~phone calls, between a bank (the service provider)
and applicants (the user) in a loan application process. We preprocess
the BPIC12 and BPIC17 event logs following setups from the
literature~\cite{bautista2012bpic12challenge,rodriguesStairwayValueMining,kobialka2025decision}. BPIC17
records a change in the underlying behavior, a so-called \emph{concept
  drift}~\cite{adams2021framework}. Thus, we split BPIC17 into BPIC17a
and BPIC17b containing traces before and after the concept drift,
respectively. BPIC12 contains $5\,053$, BPIC17a $10\,746$ and BPIC17b
$12\,344$ traces. With preprocessing, BPIC17a reduces to $5\,515$ and
BPIC17b $6\,989$ traces; BPIC12 remains at the same size.

\begin{comment}
  \red{\textbf{MSSD (5)}.} \red{The Music-Streaming Session Dataset
    (MSSD) provides listening sessions recorded by the music streaming
    service Spotify.  Brost
    \etal~\cite{brostMusicStreamingSessions2019} published the MSSD
    with six challenges to accelerate research on listening behavior.
    In \emph{Challenge 4}, Brost
    \etal~\cite{brostMusicStreamingSessions2019} suggests to analyze
    ``user journeys'' to influence the users' ``mood'". We address
    this challenge in our work.  Each listening session consists of
    10--20 tracks enriched with additional user interactions like
    listening context and pauses.  The full dataset contains 160
    million listening sessions with 3.7 million distinct tracks.  We
    use the ``mini'' version of the MSSD containing $10\,000$
    listening sessions.  }
\end{comment}

\subsection{Experimental Design and Setup}

To find adequate state representations for AL and PM for the
real-world case studies, we exploit the insights gained in the
benchmark evaluation presented in \cref{sec:syn-exp},
\rnew{and used the same parameter setup  for HYBRID.}
Note that when
applying AL and PM to logs, the techniques have different semantic
definitions of states in their learned TSs: AL defines states by the
distribution over the observed future states, while PM defines states
over a state representation function including, e.g., a trace prefix
and suffix.\looseness=-1

In practice, model performance and later analysis benefit from small
models with a feasible number of loops. For PM, we propose to generate
models with the state representations in question and choose a
possible small model with a feasible number of loops. This approach
was approximated by minimizing the sum of loops and states in
\cref{sec:syn-exp}.  For AL, low values for $\alpha$ promote state
merging, whereas a higher $\alpha$ prevents merging. Since user
journeys tend to follow sequential structures and loops should be
reduced to a necessary minimum, we use
% might prefer
a more restrained
state merging for learning user journeys than the parameter setting
proposed in the literature, as discussed in \cref{sec:syn-exp}.
% would suggest~\cite{DBLP:journals/ml/MaoCJNLN16}.

\textbf{\rnew{Learning Setup.}} For the state representation in PM, we chose state representations for
trace prefix and suffix which are
% \red{prefix representation of length 3 with a list structure} list
% of past events of size $3$ to discover a model that has
a good compromise between the number of states and the number of
loops.  For AL, we evaluate the impact of the confidence parameter
$\alpha$ by considering two setups: (1)~$\alpha$ set by
$\alpha_\mathrm{approx}$ in \cref{eq:alpha} using similar
alphabet-size coefficients as in \cref{sec:syn-exp}, and (2)~low
confidence ($\alpha = 0.9$). For \name{Hybrid}, we use the setup of \name{PM-uj} and
\name{AL-uj}, in case PM or AL is selected, respectively. We set
$\lambda = \lambda_\mathrm{approx}$ similar as in \cref{sec:syn-exp}. \rnew{To automatically decide between AL and PM in \name{Hybrid}, we compare $\lambda_\mathrm{approx} > \alpha_\mathrm{approx}$ (see \cref{sec:hybrid}).
Note that we limit the comparisons between certainty threshold $\lambda_\mathrm{approx}$ and confidence
$\alpha_\mathrm{approx}$ to two digits.}

For each case study, we consider two different event logs, with and without preprocessing:
(1)~the \emph{full event log} keeps
% the
single-occurrence
% unique
traces and does not enumerate duplicate events.  (2)~the
\emph{preprocessed event log} removes all single-occurrence
% unique
traces and enumerates iterative duplicated events,
e.g., Offer~1, $\dots$, Offer~2, etc.
In the sequel, we refer to different experimental setups by a string
stating the learning technique, an indication if the log is
preprocessed, and special algorithm parameters, e.g.,
\name{AL-full-0.9} would refer to AL on the
non-preprocessed event log with $\alpha = 0.9$, and \name{AL-uj} to
AL on the preprocessed event log with the computed
$\alpha = \alpha_\mathrm{approx}$ (cf.~\cref{eq:alpha}). The experiments were run in the
evaluation framework described in \cref{sec:machine}, using only
machine (1).

% \begin{comment}
%   \red{ For MSSD, we construct listening sessions from the tabular
%   dataset by concatenating profile, track, and interaction states.
%   Each listening journey starts with the user's profile state.
%   Then, a track state for each played track is introduced,
%   differentiated by controllable and uncontrollable context,
%   e.g. think shuffle mode or user-generated playlist.  If users take
%   a \emph{long} or \emph{short} pause between playing tracks or
%   \emph{skip} a track, a corresponding state is added.
%   Additionally, we preprocess the event log by discretizing track
%   states by four key features: \emph{acousticness},
%   \emph{danceability}, \emph{energy}, and \emph{release decade}.  In
%   the full version, each track is represented by its unique
%   identifier.  }
% \end{comment}

We do not explicitly highlight in the following plots
the individual results for \name{Hybrid} as it completely overlaps
with either \name{PM-uj} or \name{AL-uj}. \name{PM-uj} is always selected in the
GrepS case study, and \name{AL-uj} in all three BPIC case studies. This
selection corresponds to our intention to use PM for sparse logs and
AL for well-distributed logs.
% In the following, we call the DFS generated with $S^{[t]_3}$ from
% $L_1$ and $L_1^{full}$ ``PM'' and ``PM-full'', and the MCs generated
% with Alergia ``AL'' and ``AL-full'' respectively and additionally
% annotate the confidence bound in $\%$, i.e. ``AL-0.9'' or ``AL''.

\textbf{RQ2.}  To investigate the impact of sparse event logs on
learning, we split the event logs into two disjoint logs: a training
log and a test log. The training log is used to learn the TS and the
test log to evaluate the generality of the learned model.  For this
purpose, we consider training and test suites with different
proportions of the full event logs going into the training log and
into the test log.  We divide the event logs with proportions ranging
from $0.4$ to $0.9$ of the traces going into the training log, in
$0.1$ increments (thus, each increment corresponds to 10\% of the
traces in the log).  Due to the randomness in the division of logs,
ten training and test suites are created for each proportion--wise
split. We then run the traces of the test log on the learned model. We
say that a trace $\sigma$ in the test suite \emph{fails} for a
$\mathit{TS}$ if $\sigma \notin \mathcal{L}(\mathit{TS})$, otherwise
it is accepted. We consider event logs with and without
preprocessing. \looseness=-1

\textbf{RQ3.}  The results reported in \cref{sec:syn-exp} and
\cref{tab:overview-results} suggest that the compared learning
techniques yield different TSs. To compare the models, we analyze the
number of overlapping failed traces for the different TSs. For this,
we run all failed traces from TS $A$ on TS $B$ and vice
versa. However, these test-based experiments are biased towards the
\emph{generality} of models, where a general model accepts more traces
from an event log. Therefore, we also investigate the differences in
the defined languages by the learned TSs.  To quantify the overlap
between the accepted languages of the two models, we define one model
to be the ground truth and test the \emph{recall} of the other model
by creating a transition coverage test suite as described in
\cref{sec:syn-exp}.
% To quantify the overlap of accepted languages between the two
% models, we approximate their languages by sampling random traces in
% one model and testing them on the other.
If we do the same in both directions, we can see the overlap between
the languages. We assume that TS $A$ is more general than TS $B$ if
$A$ accepts many traces generated by $B$, but $B$ accepts fewer traces
generated by $A$.

As an additional metric to evaluate the generality of a TS, we use the
average frequency of traversing a transition by running all
traces. The traversal frequency of a transition is the sum of
appearances of the transition in the set of runs generated when
running all traces in an event log.

\subsection{Results}

\begin{comment}
{\scriptsize
\begin{table*}[t]
  \caption{Execution time and the number of states of the learned TS for the different AL and PM techniques.} \label{tab:overview-results}
    \centering
    \setlength{\tabcolsep}{3pt}
    \begin{tabular}{l|cc|cc|cc|cc|cc|cc}
        & \multicolumn{2}{c|}{\textbf{AL-full-0.9}} & \multicolumn{2}{c|}{\textbf{AL-full}} & \multicolumn{2}{c|}{\textbf{PM-full}} & \multicolumn{2}{c|}{\textbf{AL-0.9}} & \multicolumn{2}{c|}{\textbf{AL}} & \multicolumn{2}{c}{\textbf{PM}} \\
        & \textbf{time (s)} & \textbf{\#states} & \textbf{time (s)} & \textbf{\#states} & \textbf{time (s)} & \textbf{\#states} & \textbf{time (s)} & \textbf{\#states} & \textbf{time (s)} & \textbf{\#states} & \textbf{time (s)} & \textbf{\#states} \\ \hline 
        %\textbf{Survey Sys.} & $6.9 \cdot 10^{-4}$ & 6 & $4.8 \cdot 10^{-4}$ & 5 & $5.6 \cdot 10^{-4}$ & 8 & \multicolumn{2}{c|}{n.a.} & \multicolumn{2}{c|}{n.a.} & \multicolumn{2}{c}{n.a.} \\
        \textbf{GrepS} & $3.3 \cdot 10^{-3}$ & 37 & $1.5 \cdot 10^{-3}$ & 19 & $9.6 \cdot 10^{-4}$ & 20 & $5.8 \cdot 10^{-3}$ & 40 & $2.7 \cdot 10^{-3}$ & 28 & $1.6 \cdot 10^{-3}$ & 29\\
        \textbf{BPIC12}   & 0.12 & 84  & 0.03 & 32 & 0.10 & 384 & 0.19 & 108& 0.04 & 42 & 0.11 & 707 \\ 
        \textbf{BPIC17a} & 1.77 & 156 & 0.12 & 39 & 1.13 & 337 & 0.07 & 68 & 0.04 & 43 & 0.48 & 143 \\
        \textbf{BPIC17b} & 2.65 & 194 & 0.15 & 42 & 0.32 & 394 & 0.11 & 88 & 0.04 & 40 & 0.13 & 168 
    \end{tabular}
\end{table*}}
\end{comment}

\begin{table*}[t]
  \caption{Execution time and the number of states of the learned TSs for the different AL and PM techniques.} \label{tab:overview-results}
    \centering
    \scriptsize
    \setlength{\tabcolsep}{2pt}
    \begin{tabular}{lccccccc@{\extracolsep{10pt}}c@{\extracolsep{2pt}}cccccc}\toprule
  %      \multicolumn{1}{c}{} &\multicolumn{6}{c}{\textbf{Full log}} & \multicolumn{6}{c}{\textbf{Preprocessed log}}\\\cline{2-7} \cline{8-13}\\[-7.5pt]
        & \multicolumn{2}{c}{\textbf{AL-0.9 full log}} & \multicolumn{2}{c}{\textbf{AL full log}} & \multicolumn{2}{c}{\textbf{PM full log}} & \rnew{\textbf{Events}} & \multicolumn{2}{c}{\hspace{-2mm}\textbf{AL-0.9 preprocessed}} & \multicolumn{2}{c}{\textbf{AL preprocessed}} & \multicolumn{2}{c}{\textbf{PM preprocessed}} & \rnew{\textbf{Events}}\\
                             & time (s) & \#\,states& time (s) & \#\,states& time (s) & \#\,states & & time (s) & \#\,states & time (s) & \#\,states & time (s) & \#\,states & \\ \cmidrule{2-8}
                             \cmidrule{9-15}
% & t (in s.) & $\lvert \Gamma \rvert$ & t (in s.) & $\lvert \Gamma \rvert$ & t (in s.) & $\lvert \Gamma \rvert$ & t (in s.) & $\lvert \Gamma \rvert$ & t (in s.) & $\lvert \Gamma \rvert$ & t (in s.) & $\lvert \Gamma \rvert$ \\ \hline 
        %\textbf{Survey Sys.} & $6.9 \cdot 10^{-4}$ & 6 & $4.8 \cdot 10^{-4}$ & 5 & $5.6 \cdot 10^{-4}$ & 8 & \multicolumn{2}{c|}{n.a.} & \multicolumn{2}{c|}{n.a.} & \multicolumn{2}{c}{n.a.} \\
        \textbf{GrepS} & $4.2 \cdot 10^{-3}$ & 37 & $1.1 \cdot 10^{-3}$ & 19 & $1.4 \cdot 10^{-3}$ & 20 & 25 & $3.9 \cdot 10^{-3}$ & 40 & $1.2 \cdot 10^{-3}$ & 28 & $1.1 \cdot 10^{-3}$ & 29 & 25 \\
        \textbf{BPIC12}   & 0.11 & 84  & 0.03 & 32 & 0.12 & 384 & 18 & 0.21 & 108& 0.04 & 42 & 0.12 & 707 & 18\\ 
        \textbf{BPIC17a} & 1.79 & 156 & 0.11 & 39 & 1.09 & 337 & 23 & 0.06 & 68 & 0.03 & 43 & 0.46 & 143 & 26 \\
        \textbf{BPIC17b} & 2.67 & 194 & 0.13 & 42 & 0.31 & 394 & 23 & 0.12 & 88 & 0.04 & 40 & 0.13 & 168 & 26 \\
        \bottomrule
    \end{tabular}
\end{table*}

\begin{comment}
  %first table
  \red{\textbf{MSSD-full}} & \textbf{T.O.} & - & \textbf{T.O.} & - &  & \\
  %second table
  \textbf{\red{MSSD}} & 4474.3 & 2267 & 3459 & 2065 &  &  \\
\end{comment}

% \begin{wraptable}[10]{r}{0.4\textwidth}
\begin{table}
  \vspace{-0.59cm}
  \footnotesize
      \centering
      \caption{States and loops of DFS $S^{\{t\}_r}_{[h]_c}$ for BPIC17a for row and column indices $r$ and $c$; loops are given in parenthesis.}
      \label{tab:pm_model_comparison}
%      \scriptsize
% \vspace{-0.1cm}
      \begin{tabular}{c@{\;\;}cccc}
        \toprule
     \textbf{Prefix} & \multicolumn{4}{c}{\bf Suffix length} \\
\textbf{length} & 0 & 1 & 2 & 3 \\\midrule
1 &  27\,(103) &  79\,(46) &  156\,(5) &  265\,(0)  \\
2 &  69\,(88) &  \textbf{143\,(5)} &  248\,(0) &  380\,(0)  \\
3 &  123\,(16) &  226\,(0) &  356\,(0) &  512\,(0)  \\
4 &  186\,(0) &  311\,(0) &  467\,(0) &  632\,(0) \\\bottomrule 
      \end{tabular}
% \end{wraptable}
\end{table}

\Cref{tab:overview-results} shows the execution time and the
number of states of the learned TSs for the different learning setups
in all case studies. \rnew{The computation times are measured over one execution as the algorithms are determinstic.}
We considered an event log (without
preprocessing, \rnew{indicated by ``full''}) and a preprocessed event log.  Note that all learned
TSs define a language that includes all traces in the event logs. The
results show little execution-time differences, but Alergia's
iterative state-merging had an impact on the BPIC17 case
studies. Comparing the two AL setups, we observe that setting
$\alpha = \alpha_\mathrm{approx}$ (cf.\,\cref{eq:alpha})
yields TSs with fewer states than $\alpha = 0.9$.  The models
generated with \name{PM-uj} are mostly larger in the number of
states, which relates to the construction of DFSs.\looseness-1

To choose state representations for \name{PM-uj}, we compare different horizons
for trace projection and structures for state projection.
\Cref{tab:pm_model_comparison} presents state and loop numbers for the different trace
projections when using \emph{set} structures for prefixes and list
structures for suffixes on the BPIC17a event log. \Cref{tab:pm_model_comparison} shows that prefix representations of length $2$ and suffix
representations of length $1$ result in a reasonable
number of states and loops (cf.~parenthesis), still
maintaining sufficient generality.  We repeated this process for all case
studies, selecting individual state representations, leveraging domain
knowledge from
\citet{bautista2012bpic12challenge,rodriguesStairwayValueMining,kobialka2025decision}.\looseness=-1

\rnew{To evaluate the quality of the learned models under a lack of ground-truth models, we conduct additional experiments in \textbf{RQ2} and \textbf{RQ3}.}

% \emph{Learning under sparse event
% logs.} % different pre-processing resulting in different dataset semantic definitions on each state, i.e., the content of each state is semantically defined through the state representation function before building the model.  In practice, the semantic definition requires insights into the model generation process, e.g., when (partial) traces are stored in a list or in a set has different implications on the generated model, and the adaptation of the confidence bound requires knowledge of the possible event distributions in the underlying process.  We investigate these differences with the survey system in Example~\ref{example:survey-system}.  todo write comparison todo include reason for PM parameter settings & introduce easier model names than MC-full-09 etc. - AL / PM After preprocessing, we generate user journey models with the PM and AL approach.

% For the AL approach, we conducted all experiments with two values
% for $\alpha$: (1) $\alpha$ defined by \cref{eq:alpha}, and (2)
% $\alpha = 0.9$. To evaluate \emph{overapproximations}, we use the
% models generated by $\alpha = 0.9$ as a comparison for \name{AL-uj} and
% \name{PM-uj}, \red{as $\alpha = 0.9$ should represent an
% \emph{underapproximation}}.

\begin{figure*}[t]
  \begin{subfigure}[t]{.24\textwidth}
    \includegraphics[trim=8 8 8 8,clip,width=\linewidth]{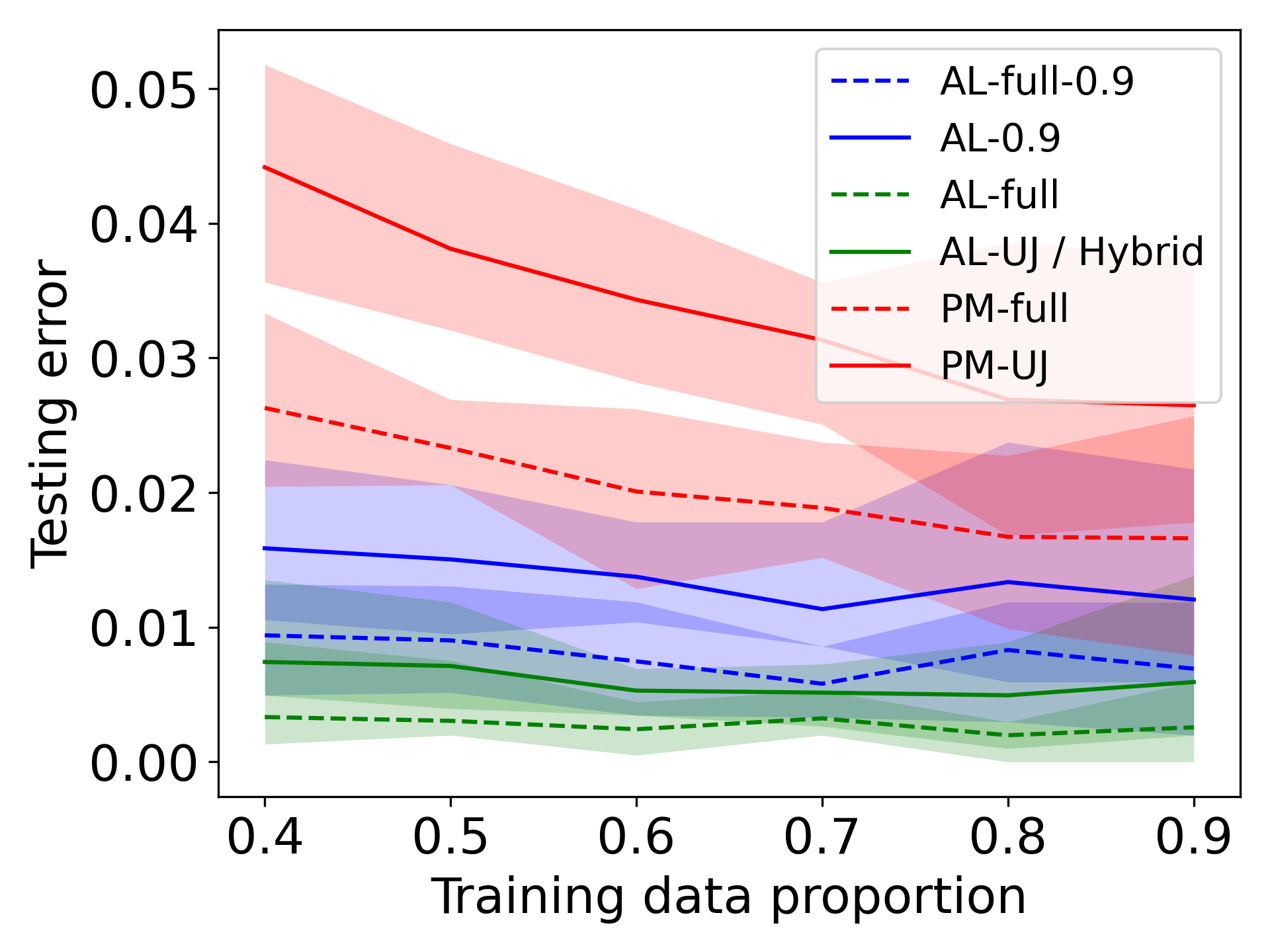}
    \caption{BPIC12.}
    \end{subfigure}
  \begin{subfigure}[t]{.24\textwidth}
    \includegraphics[trim=8 8 8 8,clip,width=\linewidth]{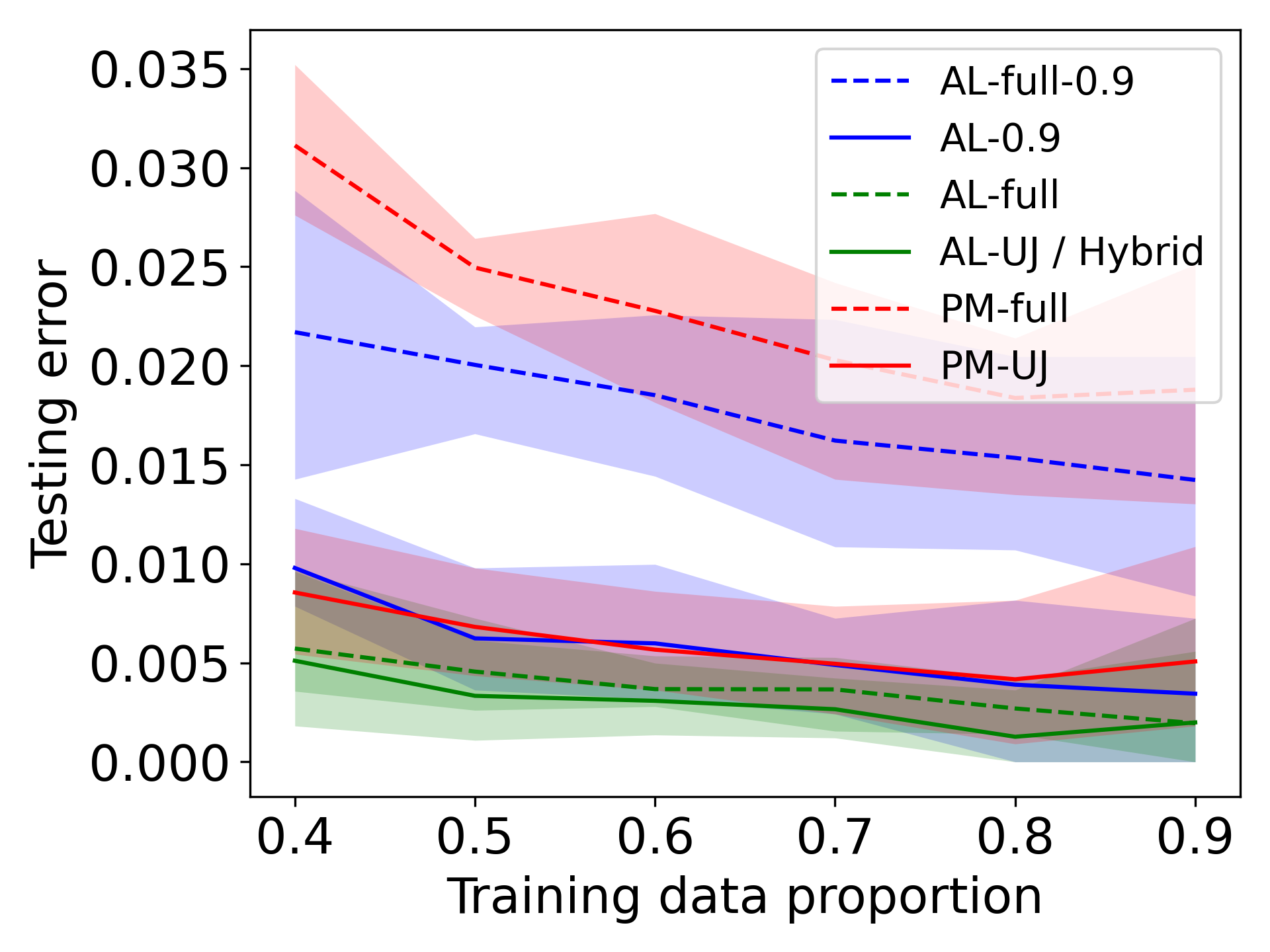}
    \caption{BPIC17a.}
    \label{fig:prob_comparison}
    \end{subfigure}
    \hfill
  \hfill
  \begin{subfigure}[t]{.24\textwidth}
  \includegraphics[trim=8 8 8 8,clip,width=\linewidth]{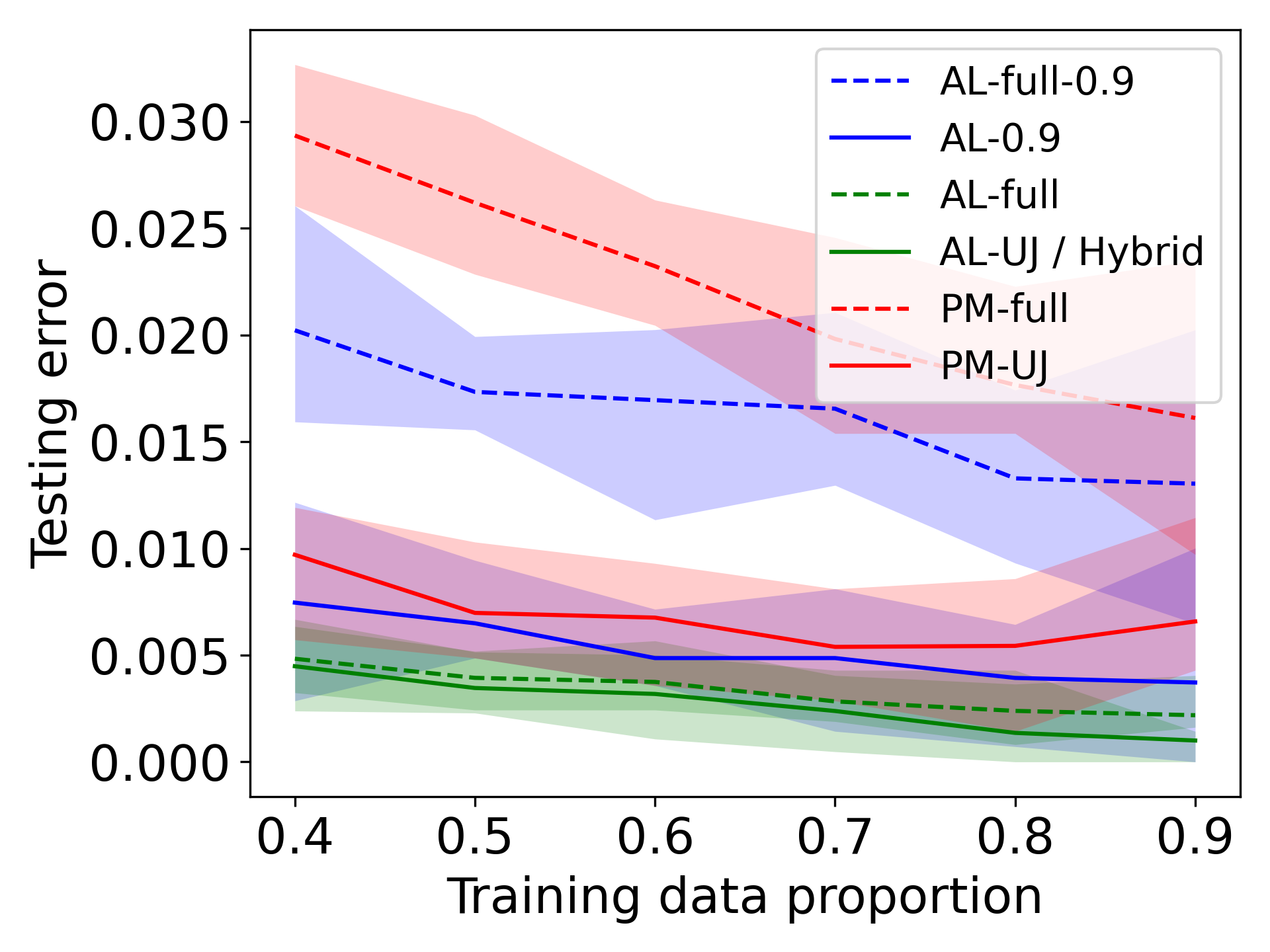}
  \caption{BPIC17b.}
  \end{subfigure}
  \hfill
  \begin{subfigure}[t]{.24\textwidth}
  \includegraphics[trim=8 8 8 8,clip,width=\linewidth]{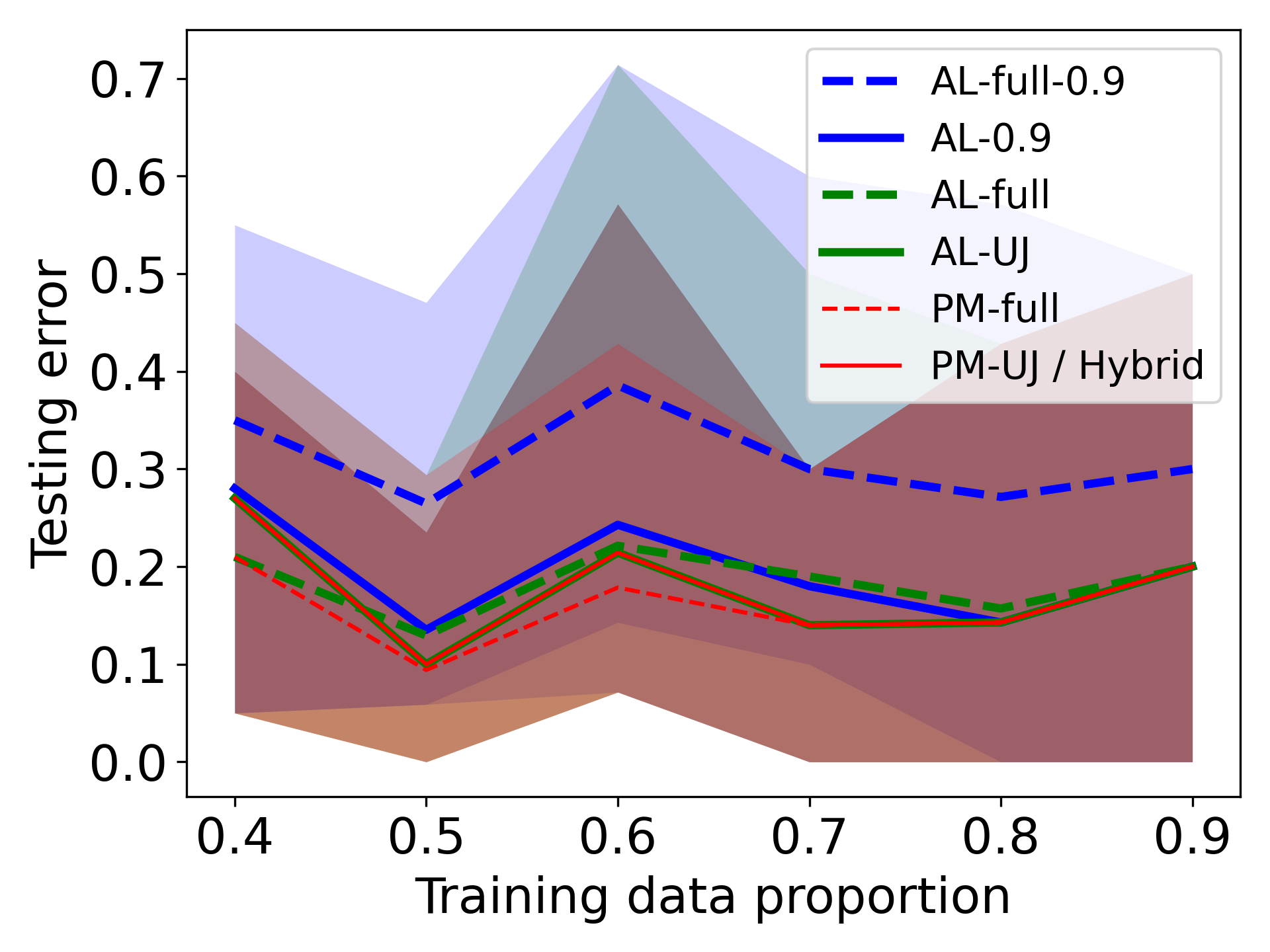}
  \caption{GrepS.}
  \end{subfigure}
  \caption{Proportion of failed test traces for the case studies.
    % with training log proportions ranging from \red{$0.4$--$0.9$}.
    \label{fig:prob_comparison_other}}
\end{figure*}

\textbf{RQ2.}
% \emph{Learning under sparse event logs}.}
To evaluate the impact of sparse logs, we conducted a growing sliding
window experiment, splitting the available traces into a training and
a test log. \Cref{fig:prob_comparison_other} displays each iteration
with the percentage of test traces that failed in the corresponding
case study; the lines represent the average over the $10$ models
learned, the surrounding hue indicates maximum and minimum results.
In general, all techniques failed fewer traces when provided with
larger training logs. The models generated with AL and an $\alpha = 0.9$ serve
as a baseline for \name{AL-uj} and \name{PM-uj} to evaluate
\emph{overapproximations}, as AL with $\alpha = 0.9$ should represent an
\emph{underapproximation}. \rnew{Overall, we see that the testing error for AL is low, which might be an indicator for an over-approximation. The results also show that PM has similar performance as AL with $\alpha = 0.9$, which means that models learned with PM tend to represent an under-approximation.} Furthermore, AL methods generate more general
models on sparse logs than the PM method. For BPIC17, the models
created by \name{AL-0.9} and \name{PM-uj} have comparable performance
with similar generalizations on unseen test traces. Since the event
log of GrepS is very small, training with only a subset bears a high
risk that the learned model misses behavior.

The results reveal that
both \name{AL-0.9} and \name{PM-uj} benefit from preprocessing. These
results highlight that suitable preprocessing enables PM to
successfully cope with sparsity. The \name{Hybrid} method correctly
selects \name{PM-uj} on smaller event logs, such as the log provided for
GrepS. On larger event logs, such as the logs for the BPIC case
studies, \name{Hybrid} selects \name{AL-uj}, which achieves better
generalizing approximations according to
\cref{fig:prob_comparison_other}. To answer \textbf{RQ2}, our
experiments show that
% we observe that
sparse event logs---such as the log for GrepS---can be challenging
for AL, while PM creates accurate models already on sparse logs,
especially if the log is preprocessed. If the sparsity of the event
logs is reduced---such as in the logs for BPIC---AL learns models
with few failing tests, where PM might not generalize well enough on
large event logs. \looseness=-1

% \begin{comment}
% \begin{figure}[t]
%     \centering
%     %\hspace{-30pt}
%     \begin{subfigure}[t]{.49\textwidth}
%       %\vspace{-75pt}
%     \includegraphics[trim=8 8 45 40,clip,width=\linewidth]{img/performance_before.png}
%       %\vspace{-5pt}
%     \caption{Fraction of failed test traces on BPIC17a over 40$\%$ to $90\%$ splits.}
%     \label{fig:perf_before}
%     \end{subfigure}%

% \hfill%    \quad
%     \begin{subfigure}[t]{.49\textwidth}
%       \includegraphics[trim=8 8 45 40,clip,width=\linewidth]{img/performance_detailed_before.png}
%       \caption{Percentage of failed test traces per model (BPIC17a)}
%       \label{fig:perf_before_detailed}
%     \end{subfigure}

%     \caption{Performance of different models.}
%     \label{fig:prob_comparison}
% \end{figure}

%     %    \quad
%     \begin{subfigure}[t]{.49\textwidth}
%       \includegraphics[trim=8 8 45 40,clip,width=\linewidth]{img/performance_after.png}
%       \caption{Average performances (BPIC17b)}
%       \label{fig:perf_after}
%     \end{subfigure}
% \hfill%    \quad
%     \begin{subfigure}[t]{.49\textwidth}
%       \includegraphics[trim=8 8 45 40,clip,width=\linewidth]{img/performance_detailed_after.png}
%       \caption{All performances (BPIC17b)}
%       \label{fig:perf_after_detailed}
%     \end{subfigure}
%     \caption{Performance of different models over k-fold set splits.}
%     \label{fig:prob_comparison}
% %\end{figure}
% \end{comment}

% \begin{table*}[t]
%   \centering
%   \notsotiny
%   \input{img/table_full_4} \input{img/table_red_4}
%   \vspace{2pt}
  
%   \vspace{-15pt}
%   \end{table*}

\begin{figure*}
  \begin{subfigure}[t]{.24\textwidth}
    \includegraphics[trim=8 8 8 8,clip,width=\linewidth]{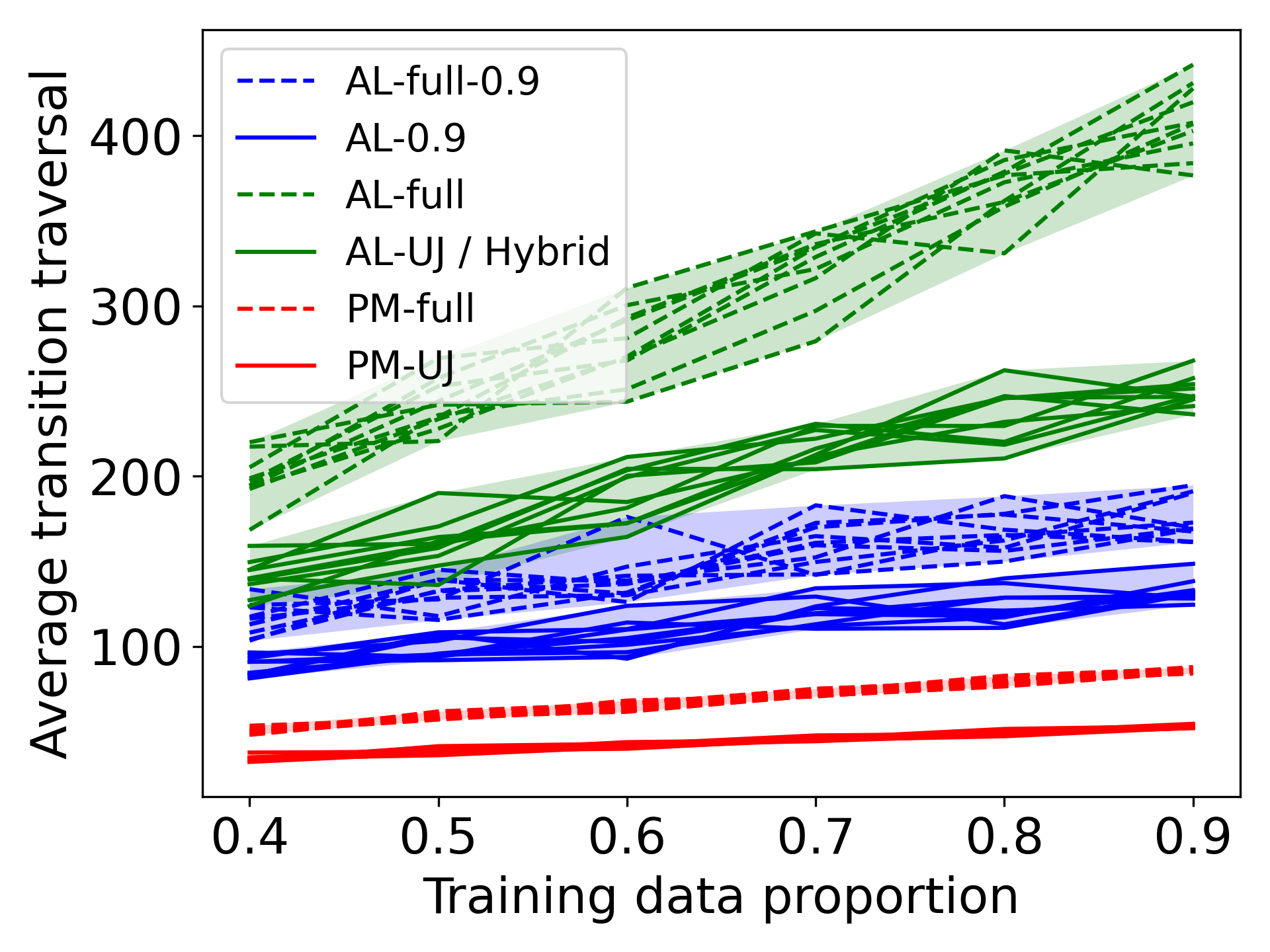}
    \caption{BPIC12.}
    \label{fig:before_average_traversal_12}
    \end{subfigure}
  \begin{subfigure}[t]{.24\textwidth}
    \includegraphics[trim=8 8 8 8,clip,width=\linewidth]{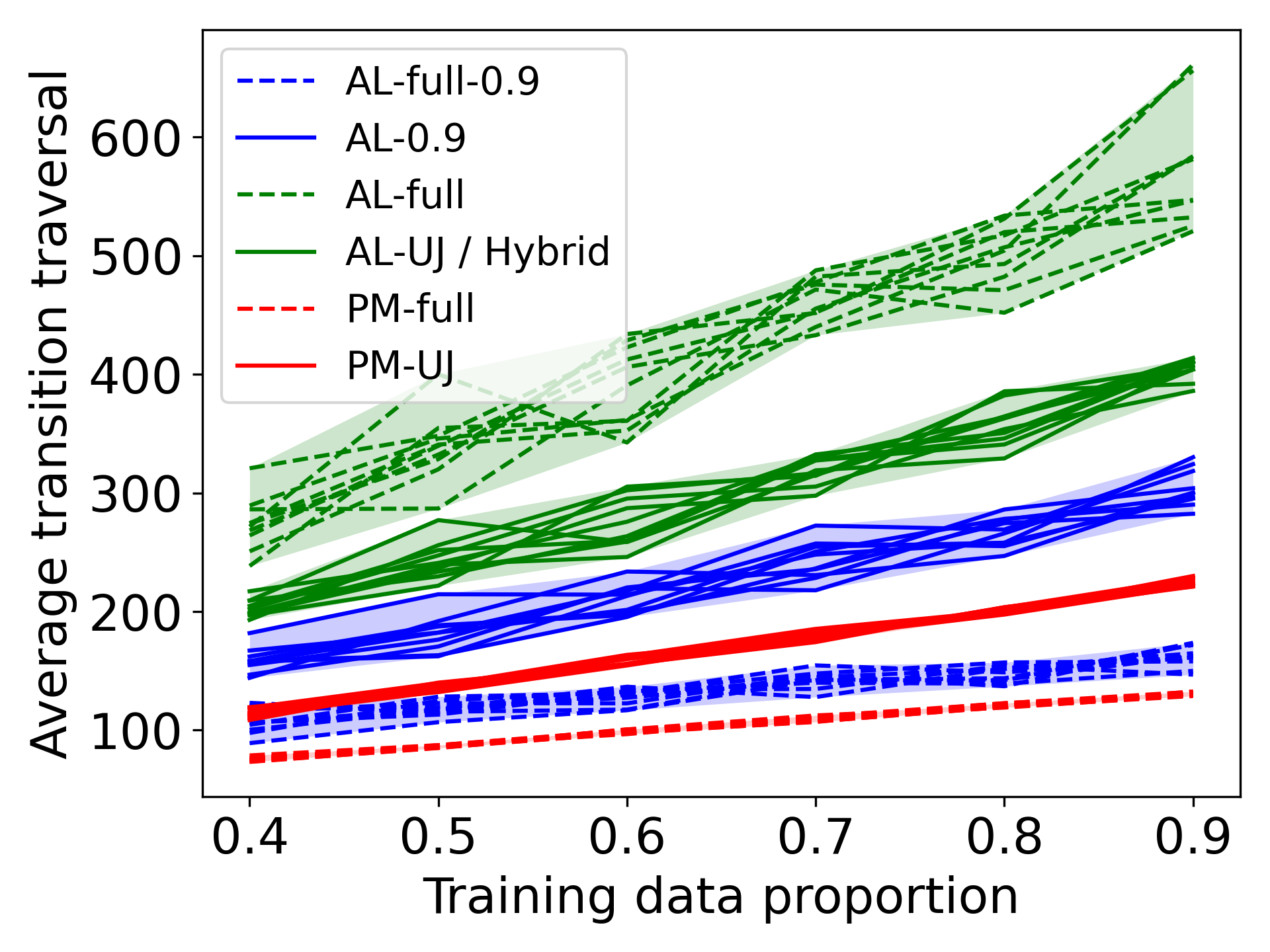}
    \caption{BPIC17a.}
    \end{subfigure}
    \hfill
  \hfill
  \begin{subfigure}[t]{.24\textwidth}
  \includegraphics[trim=8 8 8 8,clip,width=\linewidth]{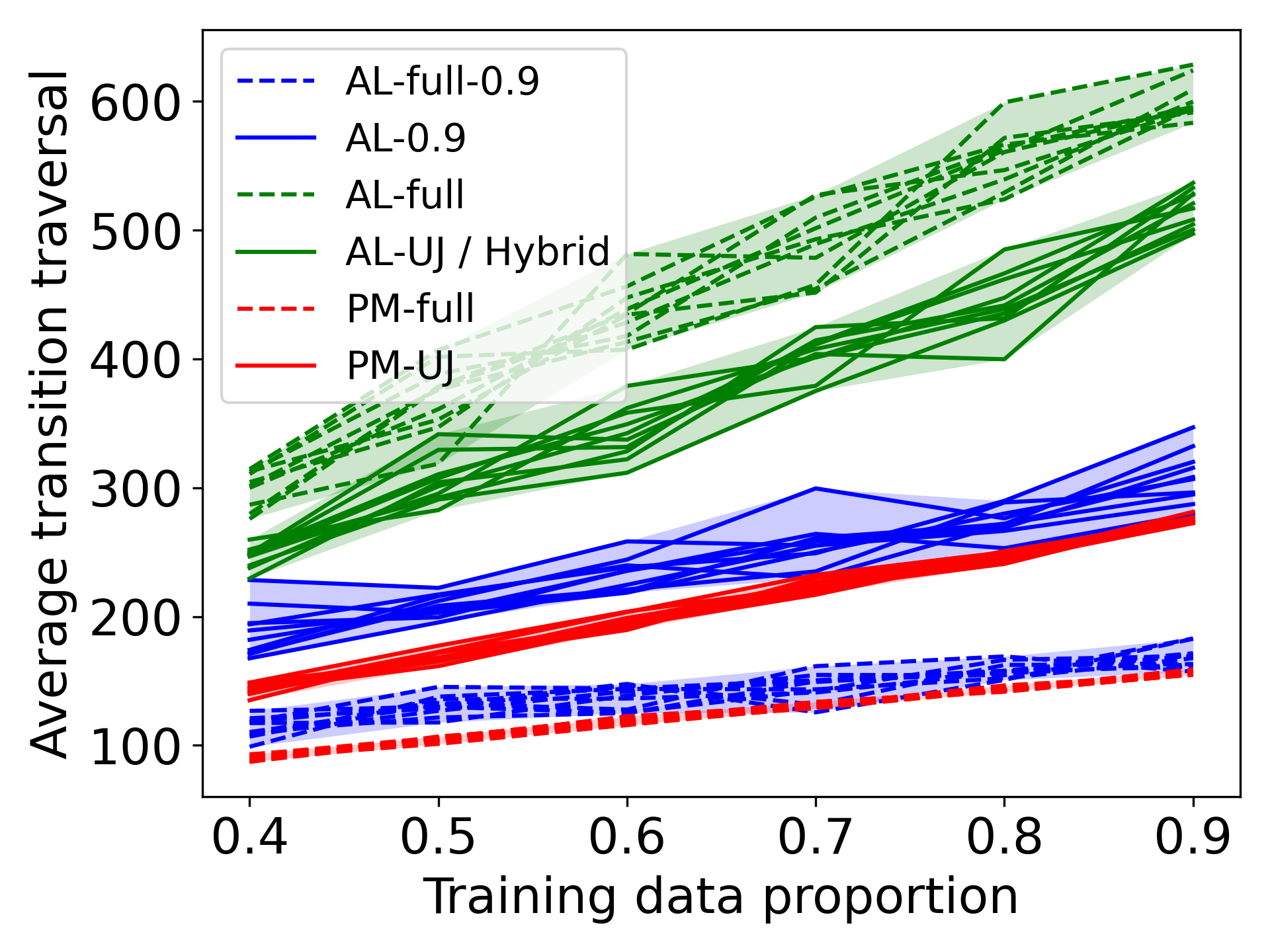}
  \caption{BPIC17b.}
  \end{subfigure}
  \hfill
  \begin{subfigure}[t]{.24\textwidth}
  \includegraphics[trim=8 8 8 8,clip,width=\linewidth]{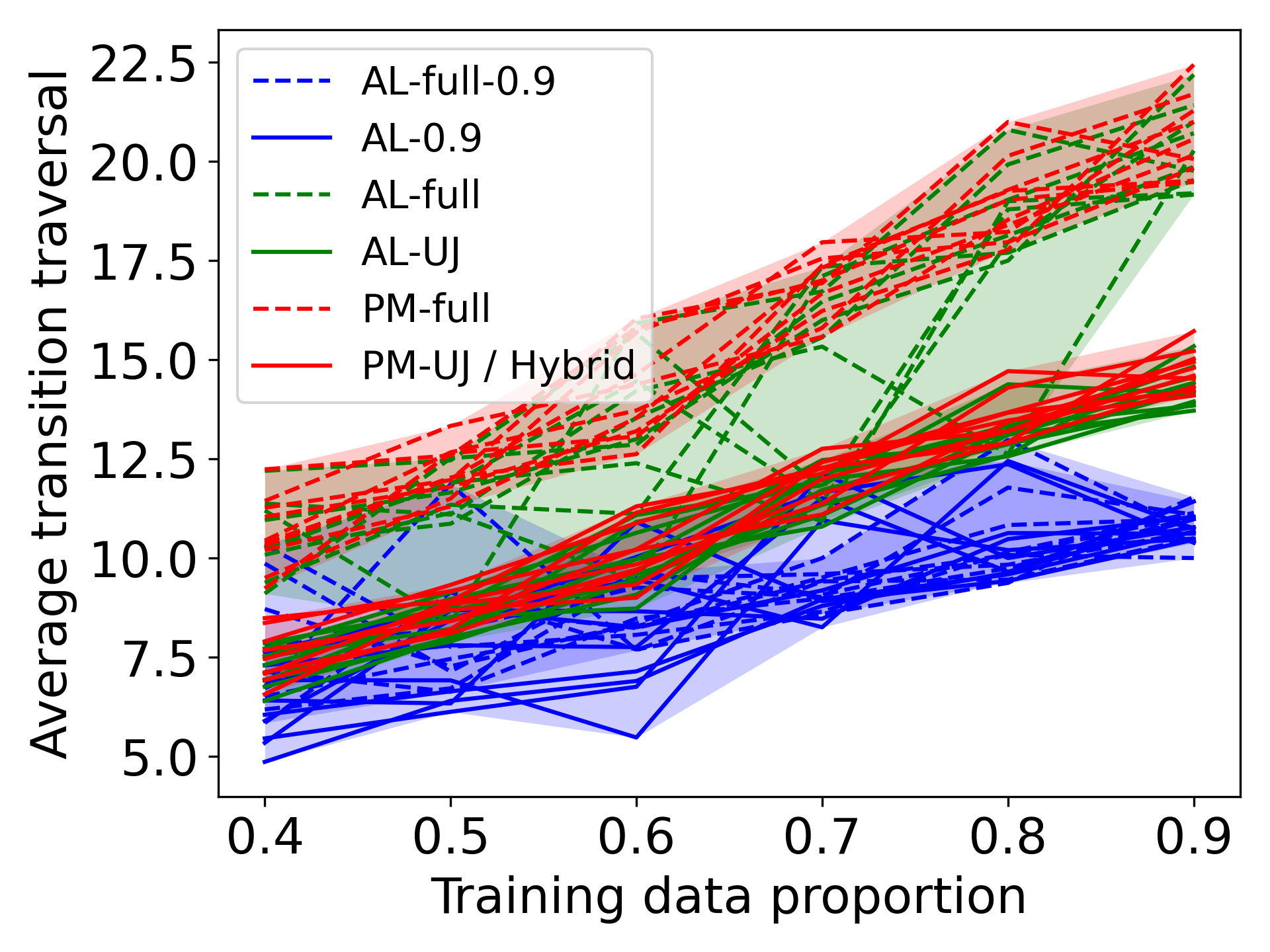}
  \caption{GrepS.}
  \label{fig:before_average_traversal_greps}
  \end{subfigure}
  \caption{Average transition frequencies for the  case studies.
    % with training log proportions ranging from $0.4$--$0.9$.
    \label{fig:transition_load_all}}
\end{figure*}

\textbf{RQ3.}
% \emph{Comparison of the learned models \textbf{(RQ2b)}.}
The results reported in \cref{sec:syn-exp} and
\cref{tab:overview-results} indicate that the evaluated techniques yield
different TSs. To compare the differences between the models, we
investigated the overlapping failing traces. \Cref{tab:table80} shows 
the average number of overlapping failing traces between models
trained on the same log; \rnew{e.g., 
%the results present on the 
for BPIC12-full, 
%case study that 
the average failing traces for \name{AL-uj} is two, we see approximately the same number of failing traces for \name{AL-0.9} and \name{PM-uj} in the same column. For \name{AL-0.9} and \name{PM-uj}, we see that $8.4$ and $16.9$ traces fail on \rnew{each} 
%the investigated 
technique, respectively. 
However, only $5.2$ failing traces overlap for \name{AL-0.9} and \name{PM-uj}.} In general, \cref{tab:table80} reports a similar trend between the
full and preprocessed logs. The results show that approximately all
failing traces of \name{AL-uj} models also fail for \name{AL-0.9} and
\name{PM-uj} models. Recall from \cref{sec:preliminaries} that a TS
$\mathit{TS}_A$ defines \emph{more general behavior} than another TS
$\mathit{TS}_B$ if
$\mathcal{L}(\mathit{TS}_B) \subseteq \mathcal{L}(\mathit{TS}_A)$
holds. This indicates that the \name{AL-uj} models represent an
overapproximation of the \name{PM-uj} and \name{AL-0.9} models. Further,
the number of overlapping tests is around half of the total number,
which suggests that there are behavioral differences between the
\name{PM-uj} and \name{AL-0.9} models.\looseness=-1

\begin{table*}[t]
  \setlength{\aboverulesep}{0pt}
  \setlength{\belowrulesep}{0pt}
  \centering 
\scriptsize
% \footnotesize
\setlength{\tabcolsep}{2pt}
\caption{Average overlapping failing traces with training log proportion $0.8$.\label{tab:table80}}
\vspace{-0.25cm}
\begin{tabular}{laaarrraaaccccaaacccaaaccc} \toprule \rule{0pt}{3ex}
  {} &  \multicolumn{3}{a}{\textbf{BPIC12-full}} &  \multicolumn{3}{c}{\textbf{BPIC17a-full}} &  \multicolumn{3}{a}{\textbf{BPIC17b-full}} &  \multicolumn{3}{c}{\textbf{GrepS-full}}  &  \multicolumn{1}{c}{} &  \multicolumn{3}{a}{\textbf{BPIC12}} &  \multicolumn{3}{c}{\textbf{BPIC17a}} &  \multicolumn{3}{a}{\textbf{BPIC17b}} &  \multicolumn{3}{c}{\textbf{GrepS}} \\[0.5ex]
  %{} &  \multicolumn{3}{r|}{\textbf{AL-full-}} &  \multicolumn{4}{r|}{\textbf{AL-full-0.9}} &  \multicolumn{4}{r|}{\textbf{PM-full}} \\
  {} & \rot{\name{AL-uj}} & \rot{\name{AL-0.9}} & \rot{\name{PM-uj}} & \rot{\name{AL-uj}} & \rot{\name{AL-0.9}} & \rot{\name{PM-uj}} & \rot{\name{AL-uj}} & \rot{\name{AL-0.9}} & \rot{\name{PM-uj}} & \rot{\name{AL-uj}} & \rot{\name{AL-0.9}} & \rot{\name{PM-uj}}  & & \multicolumn{1}{a}{\rot{\name{AL-uj}}} & \rot{\name{AL-0.9}} & \rot{\name{PM-uj}} & \rot{\name{AL-uj}} & \rot{\name{AL-0.9}} & \rot{\name{PM-uj}} & \rot{\name{AL-uj}} & \rot{\name{AL-0.9}} & \rot{\name{PM-uj}} &\rot{\name{AL-uj}} & \rot{\name{AL-0.9}} & \rot{\name{PM-uj}} \\
  \cline{2-13} \cline{15-26}\\[-7.5pt]
  \name{AL-uj} &          2.0 &         1.9 &      2.0  & 5.8 &          4.3 &      5.0 & 5.9 &          5.0 &      5.5 & 1.1 & 1.0 & 1.0 & \quad  &     5.0 &    4.5 &  5.0  & 1.4 &     1.2 & 0.8 & 1.9 &     1.6 & 1.8 & 1.0 & 1.0 & 1.0 \\
  \name{AL-0.9} &      1.9 &         8.4 &      5.2  & 4.3 &         33.0 &     12.6 & 5.0 &         32.8 &     16.1 & 1.0 & 1.9 & 1.0 & \quad  & 4.5 &   13.5 &  10.8  & 1.2 &     4.3 & 2.0 & 1.6 &     5.5 & 3.5 & 1.0 & 1.0 & 1.0\\
  \name{PM-uj} &          2.0 &         5.2 &     16.9  & 5.0 &         12.6 &     39.5 & 5.5 &         16.1 &     43.6 & 1.0 & 1.0 & 1.0 &  \quad \quad \quad & 5.0 &    10.8 & 27.2  & 0.8 &     2.0 & 4.6 & 1.8 &     3.5 & 7.6 & 1.0 & 1.0 & 1.0 \\
  \bottomrule
  \end{tabular}
  \bigskip
    \caption{Average recall over all generated model pairs with training log proportion $0.8$.\label{tab:simulation_08_table}}
    % \vspace{-0.25cm}
    % \setlength{\tabcolsep}{2pt}
    % \footnotesize
    \begin{tabular}{laaacccaaaccccaaacccaaaccc} \toprule \rule{0pt}{3ex}
      {} &  \multicolumn{3}{a}{\textbf{BPIC12-full}} &  \multicolumn{3}{c}{\textbf{BPIC17a-full}} &  \multicolumn{3}{a}{\textbf{BPIC17b-full}} &  \multicolumn{3}{c}{\textbf{GrepS-full}} &  \multicolumn{1}{c}{} &\multicolumn{3}{a}{\textbf{BPIC12}} &  \multicolumn{3}{c}{\textbf{BPIC17a}} &  \multicolumn{3}{a}{\textbf{BPIC17b}} & \multicolumn{3}{c}{\textbf{GrepS}}\\[0.5ex]
      %{} &  \multicolumn{3}{r|}{\textbf{AL-full-}} &  \multicolumn{4}{r|}{\textbf{AL-full-0.9}} &  \multicolumn{4}{r|}{\textbf{PM-full}} \\
      {\tikz[outer sep=0,diag text/.style={inner sep=0pt, font=\scriptsize},
      shorten/.style={shorten <=#1,shorten >=#1}]{%
        \node[below left, diag text, text width = 0.5cm] (gt) {\textbf{base}};
        \node[above right=7pt and 0pt, diag text] (test) {\textbf{test}};}}
        %\draw[shorten=1pt, very thin] (gt.north west|-test.north west) -- (gt.south east-|test.south east); }} 
& \rot{\name{AL-uj}} & \rot{\name{AL-0.9}} & \rot{\name{PM-uj}} &  \rot{\name{AL-uj}} & \rot{\name{AL-0.9}} & \rot{\name{PM-uj}} & \rot{\name{AL-uj}} & \rot{\name{AL-0.9}} & \rot{\name{PM-uj}} & \rot{\name{AL-uj}} & \rot{\name{AL-0.9}} & \rot{\name{PM-uj}} & & \rot{\name{AL-uj}} & \rot{\name{AL-0.9}} & \rot{\name{PM-uj}} & \rot{\name{AL-uj}} & \rot{\name{AL-0.9}} & \rot{\name{PM-uj}} & \rot{\name{AL-uj}} & \rot{\name{AL-0.9}} & \rot{\name{PM-uj}} & \rot{\name{AL-uj}} & \rot{\name{AL-0.9}} & \rot{\name{PM-uj}}\\ \cline{2-13} \cline{15-26}\\[-7.5pt]
         % \cmidrule(rr){1-1}
          \name{AL-uj} &          0.76 &         0.72 &      0.73  & 0.96 & 0.73 & 0.57 & 0.95 & 0.69 & 0.52 & 0.85 &         0.75 &     0.84 &    &      0.64 &         0.62 &      0.71  & 0.95 & 0.82 & 0.84 & 0.95 & 0.82 & 0.84 & 0.88 &    0.75 & 0.88\\
          \name{AL-0.9} &      0.90 &         0.87 &      0.81  & 0.96 & 0.88 & 0.74 & 0.96 & 0.87 & 0.72 & 0.87 &         0.83 &     0.88 &    &  0.75 &         0.75 &      0.74  & 0.92 & 0.93 & 0.85 & 0.98 & 0.95 & 0.91 & 0.89 &    0.84 & 0.89\\
          \name{PM-uj} &          0.99 &         0.96 &      0.93  & 0.94 & 0.78 & 0.92 & 0.98 & 0.84 & 0.93 & 0.84 &         0.76 &     0.84 & \quad \quad \quad    &      0.98 &         0.90 &      0.92  & 0.95 & 0.87 & 0.97 & 0.97 & 0.91 & 0.97 & 0.88 &    0.75 & 0.88\\
          \bottomrule
        \end{tabular}
    \vspace{2pt}
\end{table*}

\Cref{tab:simulation_08_table} shows the average recall values, with
the row indicating the assumed ground-truth models; i.e., these models
represent the basis for the assessment of the test verdict.  The
columns specify the tested model.  We observe that the models
generated by the same technique differ; e.g., on average the BPIC12
\name{AL-full} models accept a proportion of $0.76$ of the traces
generated by the ground-truth \name{AL-full} models. The results
further support our assumption about overapproximation: \name{AL-full}
accepts many traces generated by other models, but traces generated by
the \name{AL-full} model are more likely to fail on other models. By
preprocessing the log, AL tends to create traces that are more likely
to be in the language of the other models. The \name{AL-0.9} and
\name{PM-uj} models produced traces that were accepted by most other
models. On average proportions of $0.86$ and $0.90$ of the traces are
positive, respectively, which are the lowest recall values
observed. We assume that \name{AL-0.9} and \name{PM-uj} approach a
similar level of approximation. These results show that preprocessing
can prevent AL from overapproximating the event log and could be used
for sparse logs with a high variety.\looseness=-1
% However, the gap between these values indicates a difference between
% the models created using PM or
% AL-0.9. %The generally lower acceptance rate by PM models suggests that the PM models might be overfitting the data or that all the other models are overapproximations.  accepting their own traces above-average with $82.1\%$ and $96.3\%$ and \red{the PM-full models being $14.2\%$ more consistent in the language that the similarly generated models accept}.

%The model AL-full can replay on average $87\%$ of its traces, but no other model manages to replay more than $49\%$ on average, with even AL only replaying $32\%$.
%As expected, the replaying experiment confirms that models trained on $L_{1}^{full}$ accept a larger language than the models trained on $L_{1}$, i.e. their fractions are always larger than the ones of the regular preprocessed models.
%AL generalizes into the full dataset, its own traces can be replayed by the other AL models, with $89\%$ on average, can not well be replayed on AL-0.9 or PM, but can be replayed again by the models trained on $L_{1}^{full}$, PM-full replays $68\%$ on average.
%Furthermore, the models trained on $L_{1}^{full}$ adapt heavily to the rare behavior, they generate many traces that can only be replayed on models trained on $L_{1}^{full}$.

The analysis of our second metric on transition traversal frequency
indicates that \name{AL-uj} models define more general languages.
\Cref{fig:transition_load_all} compares the average transition
frequency when running the training log for each of the $10$ trained
models of each proportion-wise split with training log proportions
between $0.4$ and $0.9$.
%Figures~\ref{fig:before_average_traversal_12}--\ref{fig:before_average_traversal_greps} compare the average edge traversals in each training set and 
%Figures~\ref{fig:before_median_traversal_12}--\ref{fig:before_median_traversal_greps} compare the the median edge traversals.
The results show that the runs performed on the \name{AL-uj} model
traverse the same transitions more often. The average values are
increasing. This indicates that that a larger training log did not
necessarily lead to models with more states, but to more general
models.
%Comparing the two plots reveals that the medians show only a small increase with larger training sets, while the averages are strictly increasing.
%For AL-0.9 and PM increases the average edge load stronger than AL-full-0.9 and PM-full, indicating that model growth on $L_{1}^{full}$ is larger than on $L_{1}$.
%This does not hold for AL and AL-full, both have the highest edge load on average and in the median, meaning that their edges are utilized by the most traces, on average, with AL-full outperforming AL.
%Comparing the order between median and average reveals that PM-full introduces the largest number of transitions leading to a medium edge load of below 5.
% The average transition frequency confirms the findings from the previous experiments.
The average transition frequency of the \name{PM-uj} model grows at the
same rate as \name{AL-0.9}. A similar observation applies to the full
event logs. However, \name{AL-0.9} seems to create more general models
than \name{PM-uj}.  Further, the \name{AL-full} models have the highest
fluctuations in the average traversals, while the \name{PM-uj} models are
the most constant. This highlights that the behavior defined by models
learned by AL techniques highly depend on the underlying event log,
and that PM techniques are more robust to minor fluctuations in the
event log. In summary, we answer \textbf{RQ3} by observing that models
learned by AL are more general and have fewer states, while models
created with PM are less general and have more states.\looseness-1

\subsection{Threats to Validity}
To avoid a biased evaluation, we considered four case studies with
four event logs ranging in size from 33 to several thousand traces.
The case studies record digital services assuming that users do not
perform actions simultaneously. For \textbf{RQ2}, the results for
GrepS seem to not align with the BPIC results. We believe this is due
to the small event log provided for GrepS, where subsets of the event
log are likely to not include all behavioral aspects. Furthermore, we
note that the evaluation for \textbf{RQ2} favors models that define a
more general behavior.  %AL techniques have problems with
%sparse event logs, whereas our results in Sect.~\ref{sec:syn-exp}
%showed that PM might be advantageous for generating models with high
%precision in this setting.
Our experiments use a preprocessing adjusted for user journeys,
originally introduced by \citet{kobialkaEdba22}. Our results show that
appropriate preprocessing has a strong influence on the PM and AL
models.  To answer \textbf{RQ3}, we used a similar conformance-testing
technique based on transition coverage as in \cref{sec:syn-exp}. We
assume that transition coverage is sufficient to estimate the language
overlap as exhaustive testing methods were not computationally
feasible, see \cref{sec:threats_sync}. Although complimentary in size,
the real-world case studies in our experiments all exhibit a fairly
protocol-like behavior, where users aim for a final goal. We
believe this reflects the shape of most user journeys; user journeys
in which a large set of actions could be chosen in near arbitrary and
repetitive order lie outside of the conducted experiments. \looseness-1
% the sample the language of a model for approximating the language
% overlap instead of an exact representation, computing all accepted
% words up to a certain length proved to be computationally
% unfeasible.  Thus, the findings in \textbf{RQ2} are not absolute but
% only an approximation and it is subject to fluctuations, where we
% sampled $1\,000$ traces per model.

%%% Local Variables:
%%% mode: latex
%%% TeX-master: "main"
%%% End:

\section{Discussion on Actionable Insights}
\label{sec:actionable_insights}

In software engineering research, we aim for tools that support and
improve the development process. 
% \red{For service-oriented businesses, the user experience is crucial to the business success and software solutions need to developed considering the users behavior.} \apnote{software eginerering aspect removed?}
The success of service-oriented software
solutions depends on the user experience, which needs to be considered in the software engineering process.
%Many services are developed from an
%engineering perspective, but the success of service-oriented software
%solutions also depends on the user experience.
This paper compared AL
and PM, two techniques that provide insights into user behavior via
%TSs.
transition systems. % Bernhard: better not to burden the reader with acronyms here.
Our extensive evaluation
% and related work by \Citet{kobialkaWeightedGames,kobialkaFM24}
shows that both techniques are capable of accomplishing this
task. Still, the question remains which technique should be applied in
practice to a given event log?  \looseness-1

\textbf{Actionable Insights.}
The \name{Hybrid} method combines practical insights from our experiments,
%have been combined into the \name{Hybrid} method
and is recommended for learning behavioral models of user
journeys in practice: \name{Hybrid} automatically selects PM or
AL considering that (1) PM requires domain knowledge and is
well-suited for small event logs,
% in early development stages,
and (2) AL requires an adapted confidence parameter estimation and
larger logs.
% that might be available in later stages.
%\name{Hybrid} also benefited from a preprocessed event log,
%independent of the learning technology it selected.  
\name{Hybrid} should be applied with a preprocessing of the event log, independent of the learning technology it selected. 
Further, \name{Hybrid} can be instantiated with any PM and AL algorithm, leveraging hand-curated evaluation functions for the given log.
The following discussion provides further explanations for these findings.
\looseness-1
% even enable the application of AL to datasets otherwise
% computationally unfeasible.

For real-world applications, there exists no ground-truth
model. To evaluate the ability of AL and PM to adequately capture the
ground truth (\textbf{RQ1}, cf.~\cref{sec:syn-exp}), we created a synthetic benchmark
suite. The results in \cref{sec:syn-exp} suggest that PM techniques
are beneficial for sparse logs when expert domain knowledge is
available, i.e., knowledge about the interdependence of past and
future events. Our evaluation investigated different state
representations for PM. 
%We found that the state representation is crucial to learn accurate models. 
Without a suitable state representation,
%Otherwise, 
models generated by PM methods risk to not define the behavior adequately.
%underapproximate the behavior. 
Therefore, we suggest to apply PM techniques 
for the early stages of software engineering, when only few %little
user interactions are recorded. % experience is available.
AL can generate user-journey models
even from complex event logs, by taking into account the underlying distribution of
events in the log in the form of confidence parameter $\alpha$.
%The applied AL algorithm \emph{Alergia} regulates the confidence in the distribution of the event log by the parameter $\alpha$. 
Our results showed that the
approximation of $\alpha$ proposed by
\Citet{DBLP:journals/ml/MaoCJNLN16} allows too many states to be merged, %would merge too many states, which
creating overapproximating models. We proposed a
confidence-parameter approximation $\alpha_\mathrm{approx}$ that
supports the creation of models that are neither over- nor
underapproximations. 
%In general, AL benefits from large, well-distributed event logs. 
As AL benefits from large, well-distributed event logs, AL is better suited for
later stages in the software engineering process.  \looseness=-1

The results from four real-world case studies (\cref{sec:real-exp})
are consistent with the findings from the synthetic benchmark
(\cref{sec:syn-exp}). In \textbf{RQ2} (cf.~\cref{sec:real-exp}), we
investigated the learning algorithms for different levels of sparsity
in the underlying event log.  Our results show that AL
struggles for very sparse event logs, but creates accurate models on
larger event logs. We observed that PM could handle sparse event logs,
but learns underapproximations for large event logs.  In \textbf{RQ3}
(cf.~\cref{sec:real-exp}), we compared the behavioral differences
between the learned models, reinforcing the previous findings.
%generated by the investigated learning techniques.
For PM, we used domain knowledge to learn models established for the
case studies.  Still, the resulting models generalized less than
models learned by other techniques.

% compared different state representations and the corresponding number
% of states of the learned TSs. We observed that the chosen state
% representation significantly influences the number of states in the
% learned model. Unless the state representation was carefully selected,
% PM created underapproximations. Therefore, we required domain
% knowledge to find a suitable representations. 
For AL, the learned
models overapproximated the behavior.
% , i.e., the model described a too general behavior which may allow
% journeys that would not be observable in
% practice. % Alergia addresses this problem by limiting the state merging by a confidence parameter on the distribution of the dataset.  To overcome the problem of estimating the confidence parameter for AL, we propose an approximation using a sigmoid function based on the number of possible events and the size of the event log.
Using the proposed approximation $\alpha_\mathrm{approx}$, prevented AL from creating
overapproximations.  Note that $\alpha_\mathrm{approx}$ is optimized
for the goal-oriented behavior observed in user journeys,
%where user journeys follow
%sequential behavior. Since user journeys are goal-orientated, we
%expect the user to achieve the goal with minimal efforts required,
with limited repetitions of events in a trace.
\looseness=-1

Finally, user journeys have particular characteristics, which PM techniques
take into account by preprocessing the event log. Our evaluation
showed that these preprocessing techniques are also beneficial
for AL, as the learned models from preprocessed event logs achieve
often higher accuracy.  \looseness-1

\rnew{\textbf{Further Analysis Steps and Managerial Implications.} 
We briefly consider
analysis steps once a user journey model has been learned from
the event logs. Postprocessing of learned models is common in
PM to utilize manual model analyses: For example, observed
patterns are completed by introducing missing transitions, transitions
between states are merged, and states are merged into
% dedicated
groups representing, e.g., the different phases of the process~\cite{van2010process}.}
Furthermore, transition systems can be transformed into other process models, such as
Petri nets~\cite{van2010process}. \Citet{leemans2018robust} compares
techniques for generating process models from very large logs and
provides practical recommendations.  Furthermore,
\Citet{aichernig2018model} show that AL-generated
models can provide a baseline for model-based analysis and
verification such as testing or model-checking. Hence, it is important
to know whether the learned model represents an over- or
under-approximation. Finally, AL models can be refined through
counterexamples showing behavioral differences between the learned
model and the system under learning, adapting the idea of
\Citet{DBLP:conf/fm/WalkinshawDG09}.

\rnew{These
  % extended analysis results
  analyses can be exploited from a managerial perspective to assess service changes and usage trends. As \name{HYBRID} is an automated technique, realistic models of the current user journeys can be constructed in a timely fashion.
Leveraging automated analysis techniques~\cite{kobialkaFM24}, the models constructed by \name{HYBRID} can  support managerial decision-making processes when further developing a service.}

\rnew{\textbf{Log Quality.} Throughout this work, we implicitly assumed that the used logs were of sufficient quality.
However, event log quality is essential for applying process mining techniques successfully but in practise, most event logs are ``incomplete, noisy, and imprecise''~\cite{bose2013wanna}.
Quality aspects of event logs are actively researched in process mining, e.g., tthe one to five star rating proposed by the ``Process Mining Manifesto''~\cite{van2011process}. 
Preprocessing (cf.\ \cref{sec:preliminaries}) can enhance later analysis results by improving the quality of the log~\cite{marin2021event}.
Marin-Castro and Tello-Leal~\cite{marin2021event} review preprocessing techniques and highlight that there is not one preferable preprocessing technique, rather, multiple techniques are needed for addressing several quality aspects of the given event log.
Recently, the representativeness of event logs for the underlying process has been studied~\cite{pei2018estimating,kabierski2023addressing,karunaratne2024role} by formalizing the notion of  completeness for  event logs. All these works complement our proposed method.}
%%% Local Variables:
%%% TeX-master: "main"
%%% End:

\section{Conclusion}\label{sec:conclusion}
This paper presents a comprehensive evaluation of passive AL and PM
for learning user journey models, with the following main
contributions:
(1) A novel benchmark suite for generating ground-truth models of user
journeys with corresponding event logs.
(2) A comprehensive evaluation of AL and PM techniques, using the
benchmark set. The results show that both techniques can generate
% are suitable for generating
user journey models, but both have limitations that depend on the
given event log.
(3) A novel method, \name{Hybrid}, to automatically select between available AL
and PM algorithms depending on the properties of a given event log. Our
evaluation shows that \name{Hybrid} can support software engineering
processes with a very accurate model independent of the event log.
(4) An evaluation of the practical feasibility of applying these
methods to real-world user journeys. Evaluating the investigated
learning techniques based on their performance on sparse event logs,
we learn that sparse logs are not suited for AL, while PM handles them
well. Larger event logs improve AL results, while PM struggles to
create models that generalize well. Comparing the behavioral
differences between the learned models, we observe that AL learns
models that overapproximate the behavior; i.e., the models describe
behavior that is not observable in practice. However, PM creates
underapproximations. We further discuss actionable insights
for applying AL or PM to learn real-world user journeys. For AL, we
see a strong dependence on the distribution in the event
log. Therefore, the parameters of the learning algorithm must be
carefully selected to learn accurate models. For PM, domain knowledge
about the underlying service is required to find an adequate state
representation of the learned model. Both methods may benefit from
preprocessing the event log, ideally exploiting domain knowledge.
\looseness=-1

% AL models with a confidence of $0.9$ perform similarly to PM models
% with suitable preprocessing, both models generate traces that can be
% replayed by the most other models, and share unseen test traces that
% can not be replayed, \textbf{RQ2}.

\rchange{
\textbf{Further Analysis Steps and Future Work.} We briefly consider
analysis steps once a model of a user journey has been learned from
the event logs. Postprocessing of learned models is commonly used in
PM to utilize the manual model analysis: For example, observed
patterns are completed by introducing missing transitions, transitions
between states are merged, and states are merged into
% dedicated
groups representing, e.g., the different phases of the process.
Furthermore, transition systems can be transformed into other process models, such as
Petri nets~\cite{van2010process}.}{} 
% \Citet{leemans2018robust} compares
% techniques for generating process models from very large logs and
% provides practical recommendations.  Furthermore,
% \Citet{aichernig2018model} show that AL-generated
% models can provide a baseline for model-based analysis and
% verification such as testing or model-checking. Hence, it is important
% to know whether the learned model represents an over- or
% underapproximation. Finally, AL models can be refined through
% counterexamples showing behavioral differences between the learned
% model and the system under learning, adapting the idea of
% \Citet{DBLP:conf/fm/WalkinshawDG09}.

% AL and PM use different postprocessing, AL uses model checking to
% verify known properties of the target systems, and PM adjusts the
% learned model for further, interpretable
\textbf{Future Work.} 
For future work, we plan to investigate the use of different
model-based verification techniques, e.g. model-checking, to analyze
user behavior based on the learned models. Such automated analyses may
create valuable insights into the service design and user experience
for improving services. Furthermore, it is interesting to investigate
how models learned with AL can be integrated with established process
analysis steps used in PM. \rnew{This might also include the consideration of other case studies from PM beyond user journeys such as software product lines, e.g., as presented by Damasceno~\etal~\cite{damasceno2019learning}.}
% , \red{extending the work of \citet{kobialkaFM24}.}

% Therefore, the model generation method needs to be adjusted to
% remove detected counterexamples from the accepted traces, which
% could be achieved by removing transitions or changing the state
% representation function.

\section{Data Availability} An artifact to replicate the presented
results is  available at
\url{https://figshare.com/s/3acae9725dda09cce811}.

%%% Local Variables:
%%% mode: latex
%%% TeX-master: "main"
%%% End:

%\input{07_appendix}

% \balance
\bibliographystyle{IEEEtranN} %unsrtnat
\bibliography{references}

% \begin{IEEEbiographynophoto}{Paul Kobialka, Andrea Pferscher, Einar Broch Johnsen, Silvia Lizeth Tapia Tarifa}
% Are with the University of Oslo, Norway (email: paulkob@ifi.uio.no, andreapf@ifi.uio.no, einarj@ifi.uio.no, sltarifa@ifi.uio.no).
% \end{IEEEbiographynophoto}

% to include picture [{\includegraphics[width=1in,height=1.25in,clip,keepaspectratio]{paul}}]
\begin{IEEEbiography}[{\includegraphics[width=1in,height=1.25in,clip,keepaspectratio]{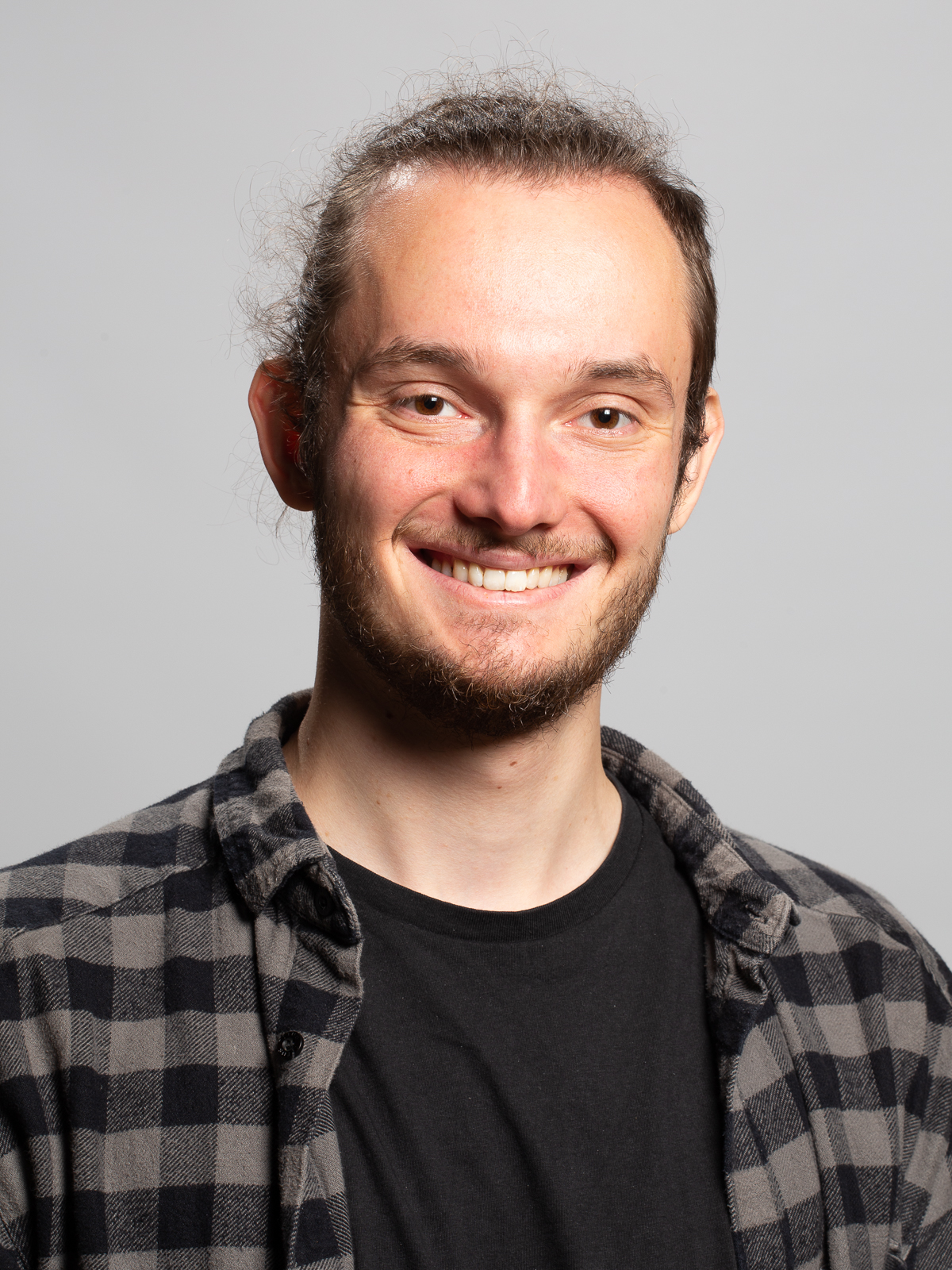}}]{Paul Kobialka} is a postdoctoral researcher in the Reliable Systems group at the Department of Informatics at the University of Oslo, where he received his Ph.D. in 2025.
Previously, he finished his Master degree in Computer Science at
RWTH Aachen University in 2021.
He is researching data-driven formal methods, i.e. problems where formal guarantees are necessary but
formal models are too complex
to construct traditionally. In his
current work, he applies formal
methods to improve user experiences in digital services 
(email: paulkob@ifi.uio.no).
\end{IEEEbiography}

\begin{IEEEbiography}[{\includegraphics[width=1in,height=1.25in,clip,keepaspectratio]{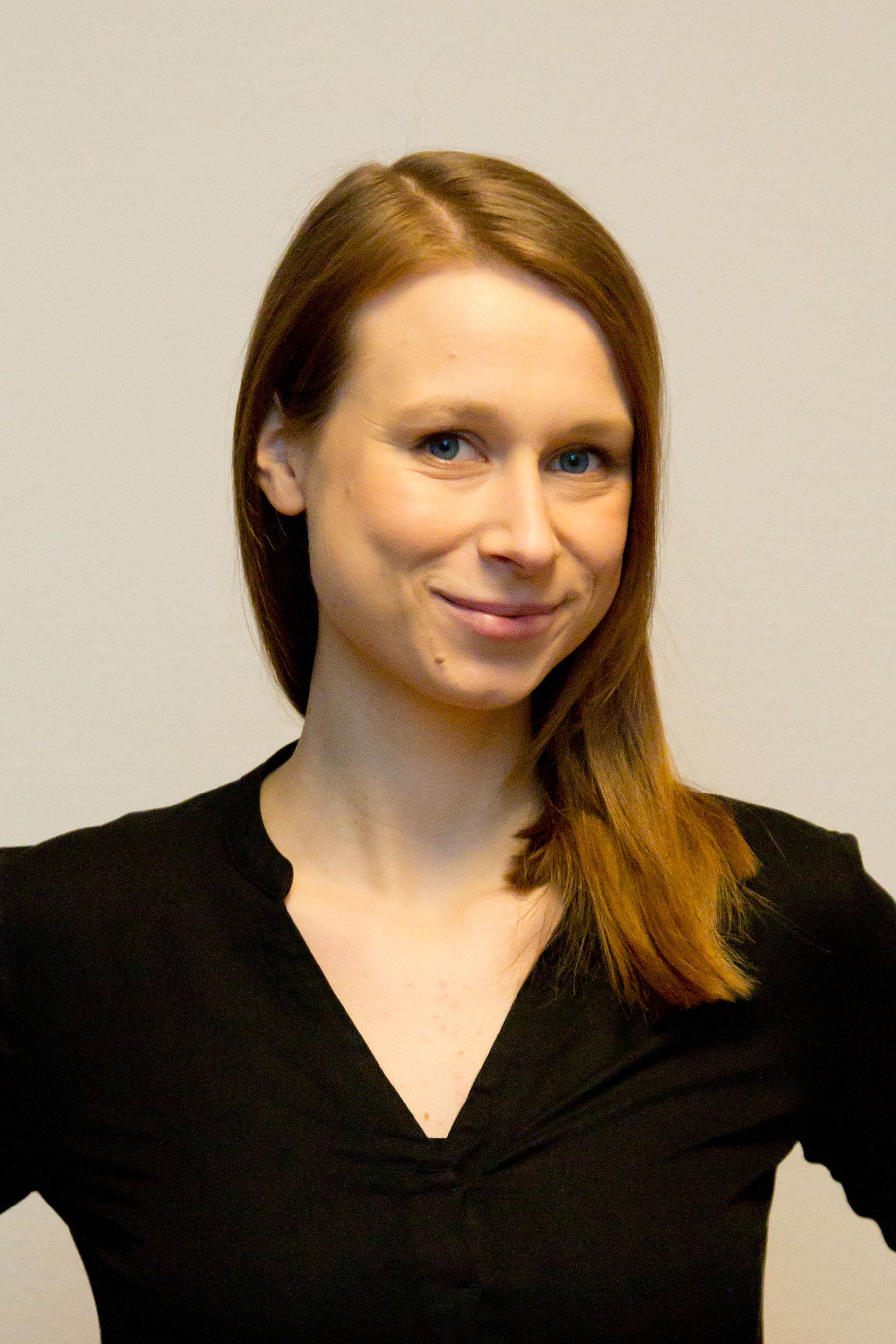}}]{Andrea Pferscher}
 is a postdoctoral researcher at the Department of Informatics at the University of Oslo. She received her PhD from Graz University of Technology in 2023. In her PhD thesis, she investigated automata learning techniques for analyzing and testing the safety-critical behavior of network components. In her current position, she has broadened her focus to the application of formal methods for digital twins on health and environmental systems. She has been a member of different Austrian, European and Norwegian research projects.
(email: andreapf@ifi.uio.no).
\end{IEEEbiography}

\begin{IEEEbiography}[{\includegraphics[width=1in,height=1.25in,clip,keepaspectratio]{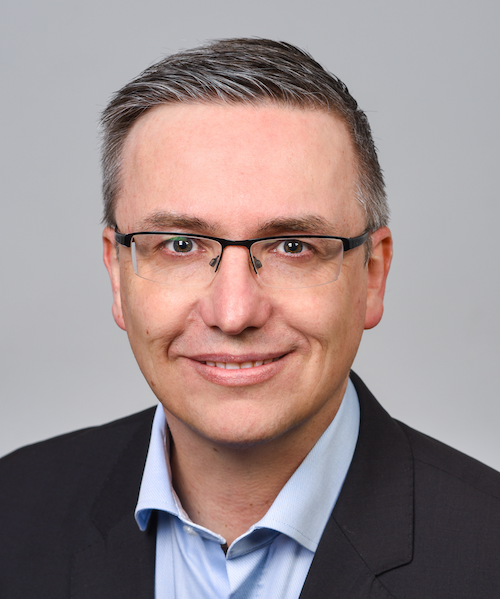}}]{Bernhard K.~Aichernig} is a full professor of Formal Methods at Johannes Kepler University Linz (JKU), Austria, where he leads the Institute of Formal Models and Verification (FMV). His research focuses on the foundations of software engineering for dependable and trustworthy systems, with interests in automated falsification, verification, and modelling. Current topics include automata learning, learning-based testing, and the integration of symbolic and subsymbolic AI. He is the author of more than 140 scientific publications.
  Until April 2025, he was affiliated with Graz University of Technology. From 2002 to 2006, he held a faculty position at the United Nations University in Macao, China. He served on the board of Formal Methods Europe from 2004 to 2016. Prof. Aichernig holds a habilitation in Practical Computer Science and Formal Methods, a doctorate, and a Diplom-Ingenieur degree, all from Graz University of Technology.
(email: bernhard.aichernig@jku.at).
%Is with the Technical University of Graz, Austria
%(email: bernhard.aichernig@tugraz.at).
\end{IEEEbiography}

\begin{IEEEbiography}[{\includegraphics[width=1in,height=1.25in,trim={5pt 0 3pt 0 },clip,keepaspectratio]{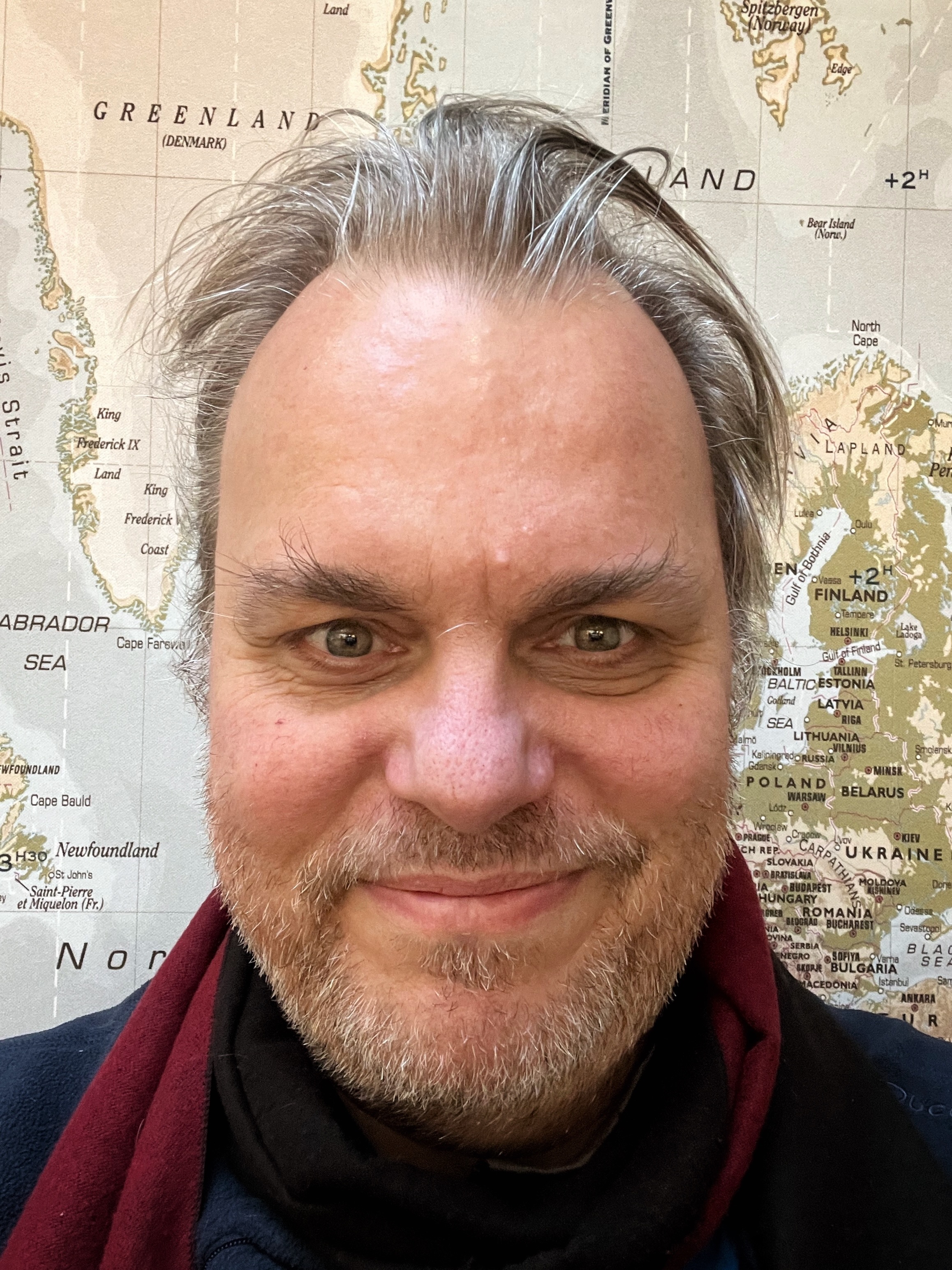}}]{Einar Broch Johnsen}
  is a professor at the Department of Informatics, University of Oslo,
  where he leads the Reliable Systems research group. He is active in
  formal methods for distributed and concurrent systems, including
  object-oriented and actor languages, manycore computing, cloud
  computing and digital twins. He is one of the main developers of the
  ABS modeling language. He has been prominently involved in many
  national and European research projects; in particular, he was the
  strategy director of Sirius, a center for research-driven innovation
  on scalable data access, he was the coordinator of the EU FP7
  project Envisage (2013–2016) on formal methods for cloud computing
  and the scientific coordinator of the EU H2020 project HyVar
  (2015–2018) on hybrid variability systems. Einar Broch Johnsen is
  member of IFIP WG2.2 “Formal Description of Programming
  Concepts”. He was board member of Sintef ICT (2009–2015). He is
  currently member of the Scientific Council of the dScience centre at
  University of Oslo, board member of Formal Methods Europe, editorial
  board member of the journals Formal Aspects of Computing and Journal
  of Logical and Algebraic Methods in Programming, and steering
  committee member of several conference series (email:
  einarj@ifi.uio.no).
\end{IEEEbiography}

\begin{IEEEbiography}[{\includegraphics[width=1in,height=1.25in,clip,keepaspectratio]{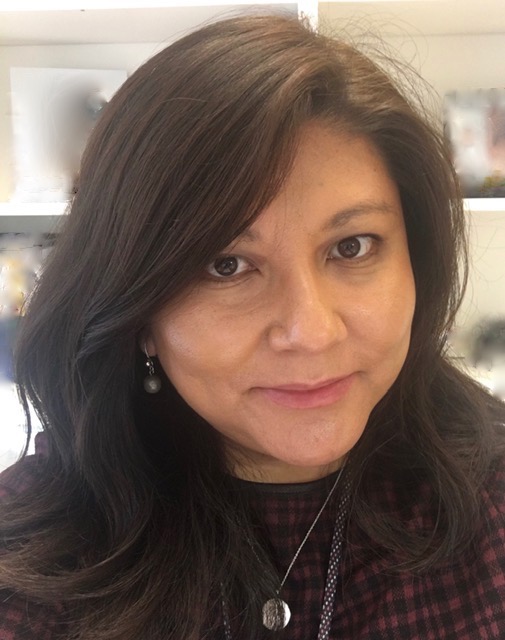}}]{Silvia Lizeth Tapia Tarifa}
  (Ph.D.  2014) is an associate professor at the Department of
  Informatics, University of Oslo. Her research activities are rooted
  in the application and foundations of formal methods, which touch on
  the study, understanding and development of formal methods, focusing
  on abstractions that benefit the specification, design and
  implementation of complex systems, the development of model-centric
  methods for distributed and component-based systems, the use of
  model-centric methods in various problem domains and the combination
  of formal methods with various existing methods, e.g., semantic
  technology, process mining, etc. In 2017, she won a young research
  talent grant, awarded by the Research Council of Norway. She has
  been contributing to several EU and Norwegian research projects
  (email: sltarifa@ifi.uio.no).
\end{IEEEbiography}

\end{document}